\documentclass[longauth]{aa}  

\usepackage{graphicx}
\usepackage{txfonts}
\usepackage{gensymb}
\usepackage{multirow}
\usepackage{graphicx}
\usepackage{times}
\usepackage{float}
\usepackage{pdflscape}
\usepackage{threeparttable}
\usepackage{amsmath}
\usepackage{amssymb}
\usepackage{xspace}
\usepackage{caption}
\usepackage{subcaption}
\usepackage{newtxtext}
\usepackage{color}
\usepackage{lscape}
\usepackage{tabularx}
\usepackage{hyperref}
\usepackage{cleveref}
\usepackage{microtype}

\usepackage{url}  

		\newcommand{\Msun}{\mbox{$M_{\odot}$}\xspace}

		\newcommand{\ergs}{\,\rm erg\,s^{-1}\xspace}
		\newcommand{\cmsq}{{\,\rm cm^{-2}}\xspace}

		\newcommand{\fexxvabs}{Fe\,\textsc{xxv}\,He$\alpha$\xspace}
		
		\newcommand{\fexxviabs}{Fe\,\textsc{xxvi}\,Ly$\alpha$\xspace}

		\newcommand{\vout}{v_{\rm out}\xspace}
		\newcommand{\mout}{\dot M_{\rm out}}

		\newcommand{\ecut}{E_{\rm cut}}

		\newcommand{\nhgal}{N_{\rm H}^{\rm Gal}}
		\newcommand{\nh}{N_{\rm H}}
		
		\newcommand{\logxi}{\log(\xi/\rm{erg\,cm\,s}^{-1})}

		\newcommand{\lbol}{L_{\rm bol}}

        \newcommand{\ekin}{\dot E_{\rm kin}}

		\newcommand{\eddratio}{\lambda_{\rm Edd}}
		
		\newcommand{\lion}{L_{\rm ion}}

		\newcommand{\mbh}{M_{\rm BH}}

		\newcommand{\macc}{\dot M_{\rm acc}\xspace}

		\newcommand{\cstat}{\mathcal{C-}\rm stat\xspace}

		\newcommand{\nustar}{\emph{NuSTAR}\xspace}		
		\newcommand{\xmm}{\emph{XMM-Newton}\xspace}

		\newcommand{\xstar}{\textsc{xstar}\xspace}
		\newcommand{\xspec}{\textsc{xspec}\xspace}

\hypersetup{
    colorlinks=true,
    linkcolor=blue,
    filecolor=magenta,      
    urlcolor=blue,
    citecolor=blue
}

\defcitealias{Tombesi15}{T15}

\begin{document}

   \title{The WISSHFUL program: \\
   the highest redshift UFO discovered in a non-lensed QSO}

\titlerunning{WISSHFUL first results}
\authorrunning{G. Lanzuisi et al. 2025}

   \author{G.~Lanzuisi\inst{1},
          L.~Borrelli\inst{2,1},
          E.~Piconcelli\inst{3},
          G.~Chartas\inst{4},
          A.~Luminari\inst{3},
          J.~Reeves\inst{5,6},
          V.~Braito\inst{5,6,7},
          E.~Bertola\inst{8},
          S.~Bianchi\inst{9},
          A.~Comastri\inst{1},
          M.~Brusa\inst{2,1},
          C.~Vignali\inst{2,1},
          F.~Vito\inst{1},
          S.~Marchesi\inst{2,1,10},
          M.~Cappi\inst{1},
          M.~Dadina\inst{1},
          L.~Zappacosta\inst{3},
          A.~Tortosa\inst{3},
          M.~Bischetti\inst{11},
          G.~Vietri\inst{12},
          F.~Salvestrini\inst{11,13},
          G.~Bruni\inst{14},
          M.~Fanelli\inst{14,15},
          E. Kammoun\inst{16},
          X. Zhao\inst{16},
          G.~Matzeu\inst{17},
          F.~Tombesi\inst{3,18,19},
          A.~Marinucci\inst{20},
          M.~Gaspari\inst{21},
          T.~Misawa\inst{22}, and         
          D.~Stern\inst{16}
          }

\institute{INAF -- Osservatorio di Astrofisica e Scienza dello Spazio di Bologna, Via Gobetti, 93/3, I-40129 Bologna, Italy, \email{giorgio.lanzuisi@inaf.it}
\and
Department of Physics and Astronomy (DIFA), University of Bologna, Via Gobetti, 93/2, I-40129 Bologna, Italy
\and
INAF -- Osservatorio Astronomico di Roma, Via Frascati 33, 00078, Monte Porzio Catone (Roma), Italy
\and
Department of Physics and Astronomy, College of Charleston, Charleston, SC, 29424, USA
\and
Institute for Astrophysics and Computational Sciences, Department of Physics, The Catholic University of America, Washington, DC 20064, USA
\and
INAF, Osservatorio Astronomico di Brera, Via Brera 20, I-20121 Milano, Italy
\and
Dipartimento di Fisica, Universit`a di Trento, Via Sommarive 14, Trento 38123, Italy
\and
INAF - Osservatorio Astrofisico di Arcetri, Largo E. Fermi 5, I-50125, Florence, Italy
\and
Dipartimento di Matematica e Fisica, Universit\`{a} degli Studi Roma Tre, Via della Vasca Navale 84, I-00146, Roma, Italy
\and
Department of Physics and Astronomy, Clemson University, Kinard Lab of Physics, Clemson, SC 29634, USA
\and
INAF – Osservatorio Astronomico di Trieste, Via G.B. Tiepolo, 11, I-34143 Trieste, Italy
\and
INAF – Istituto di Astrofisica Spaziale e Fisica cosmica Milano, Via Alfonso Corti 12, 20133 Milano,
Italy
\and
-IFPU, Institute for Fundamental Physics of the Universe, Via Beirut 2, 34014 Trieste, Italy
\and
INAF – Istituto di Astrofisica e Planetologia Spaziali, Via Fosso del Cavaliere 100, 00133 Roma, Italy
\and
Dipartimento di Fisica, Sapienza Universita di Roma, Piazzale Aldo Moro 5, 00185 Rome, Italy   
\and
Cahill Center for Astronomy \& Astrophysics, California Institute of Technology, 1216 East California Boulevard, Pasadena, CA 91125, USA
\and
Quasar Science Resources SL for ESA, European Space Astronomy Centre (ESAC), Science Operations Department, 28692, Villanueva de la Ca\~{n}ada, Madrid, Spain
\and
Department of Physics, University of Rome ‘Tor Vergata’, Via della Ricerca Scientifica 1, I-00133 Rome, Italy
\and
INFN – Sezione di Roma “Tor Vergata”, Via della Ricerca Scientifica 1, I-00133 Roma, Italy
\and
ASI – Agenzia Spaziale Italiana, Via del Politecnico snc, 00133 Rome, Italy
\and
Department of Physics, Informatics and Mathematics, University of Modena and Reggio Emilia, 41125 Modena, Italy
\and
School of General Education, Shinshu University, 3-1-1 Asahi, Matsumoto, Nagano 390-8621, Japan
}             

   \date{Received 1/04/2026; accepted -/-/-}

\abstract
{
We present the first results from the WISSHFUL program, an \xmm heritage program targeting luminous QSOs at Cosmic Noon. We report on recent simultaneous \xmm and \nustar observations of the Super-Eddington accreting quasar WISSH13 at z=3.294, which provide the highest quality broadband X-ray spectrum to date for a non-lensed QSO at this redshift. Physical modeling of the continuum reveals a soft photon index ($\Gamma\sim2$) and strong reflection ($R\sim1.4-1.8$), despite the weak narrow Fe emission, and a low high-energy cut-off ($\ecut\sim60-80$ keV, $kT_e=15-20$ keV, depending on the model adopted).
Most notably, we detect two significant (at $96.7\%$ and $98.9\%$ confidence level, respectively) absorption features at $\sim7.5$ and $\sim10$ keV rest-frame, interpreted as a blueshifted blend of \fexxvabs and \fexxviabs. These features indicate the presence of two kinematic components of a highly ionized, high column Ultra-Fast Outflow (UFO) with a velocity of $\vout\sim0.1c$ and $\vout\sim0.3c$, respectively. The slower wind is consistently detected in an archival 2017 \xmm observation, whereas the faster wind is detected only in 2024.
This stratified and variable wind exhibits extreme energetics, with a mass outflow rate of $\mout\sim20 \Msun/yr$ (corresponding to $15\%~ \macc$) for each component, and a kinetic power of the order of $\sim1$ and $\sim10\%$ of the bolometric luminosity, respectively. While this represents one of the most powerful UFOs ever detected, its kinetic power is a similar fraction of the QSO's bolometric luminosity compared to lower-redshift AGN.
We present several theoretical frameworks to explain the peculiar accretion and ejection properties of this remarkable QSO at Cosmic Noon.

}


   \maketitle

\section{Introduction}
\label{sec:intro}

\begin{table*}[t]
\begin{center}
\caption{WISSH13 multi-wavelength properties.}
\label{tab:one}
\begin{tabular}{lccccccccccc}
\hline \noalign {\smallskip}
 ID    &SDSS-ID  & z   & Log($\lbol$)  & Log($\mbh$)  & $\eddratio$ & $\macc$ &   $v^{max}_{CIV}$ & $\dot{M}^{out}_{CIV}$ & $v^{max}_{[OIII]}$ & $\dot{M}^{out}_{[OIII]}$ \\
& &  &  $\ergs$ & $\Msun$ & & $\Msun/yr$ & (km/s) & $\Msun/yr$ & (km/s) & $\Msun/yr$  \\
(1) & (2) & (3) & (4) & (5) & (6) & (7) & (8) & (9) & (10) & (11)  \\
\hline \noalign {\smallskip}
WISSH13 & J0900+4215 & 3.294  & $47.9$ & $9.3$ & 3.2 & 140 & $2580_{-160}^{+130}$ & 3 & $2380\pm180$ & 3300 {\smallskip} \\
\hline
\vspace{-0.7cm}
\end{tabular}
\end{center}
\tablefoot{
(1) WISSH ID; (2) SDSS ID; (3) Redshift; (4) Bolometric luminosity from SED fitting in \citet{Saccheo23}; (5) BH mass from H$\beta$ and (6) Eddington ratio and (7) corresponding mass accretion rate from \cite{Bischetti17}; (8) CIV maximum outflow velocity and (9) Mass outflow rate from \cite{Vietri18}; (10) [OIII] maximum outflow velocity and (11) Mass outflow rate from \cite{Bischetti17}.}
\end{table*}

Multiphase and multiscale winds driven by Active Galactic Nuclei (AGN) are thought to play a fundamental role in shaping the Supermassive Black Hole (SMBHs)/galaxy coevolution \citep{KingPounds15, Laha21} by removing and/or heating the cold gas from the host galaxy, quenching the growth of both the SMBH and the stellar component \citep{Hopkins2006} and possibly explaining the tight SMBH-host mass relation \citep[e.g.,][]{Silk98, Magorrian98}. 
To be relevant for AGN/galaxy coevolution, AGN feedback must be most effective when both BH growth and star-formation rates are at their peaks, i.e., $z\sim2-4$ \citep{Madau14, Aird15}.

Indeed, luminous QSOs at Cosmic Noon ($z=2-4$) exhibit significant evidence of powerful outflows, as highlighted by the WISSH quasar survey \citep[e.g.,][]{Bischetti17,Vietri22}. The WISSH sample consists of 85 type-1 hyper-luminous quasars with $47.2\leq Log(\lbol/[erg\,s^{-1}]) \leq 48.2$, accreting at $-1.5\leq\log(\eddratio)\leq0.5$, at $1.8<z < 4.7$, excluding lensed and radio-loud sources. A large fraction of WISSH sources show broad, blueshifted CIV emission with outflow velocities of several thousand km/s \citep{Vietri18}, a high incidence of Broad Absorption Lines in the UV \citep{Bruni19}, and [OIII] mass outflow rates among the highest reported in the literature \citep{Bischetti17}, in some cases spatially resolved (see also Bertola et al. in prep.).

The nuclear drivers of the large-scale outflows observed in different bands are thought to be winds with velocities exceeding $0.1c$, launched from the vicinity of the SMBH at inner accretion disk scales, also referred to as Ultra-Fast Outflows (UFOs, \citealt{Tombesi10}). By depositing substantial amounts of momentum and energy at the center of the host galaxy, these fast winds should be able to impact the host ISM, producing shocks that expand to large scales as energy-driven winds, with a mass loading factor in theory large enough to completely remove the host gas reservoir \citep{Zubova12,Fabian12,Fauchere12,Costa14}.

These nuclear, fast winds can be seen in the X-rays, as the highly ionized gas of which they are composed, only imprints absorption features of highly ionized elements, the most prominent ones being FeXXV/XXVI at $6.7-7$ keV rest-frame \citep[e.g.,][]{Tombesi11}. These features are now routinely observed in 30-40\% of local Seyferts and low-z QSOs \citep{Gofford13,Matzeu23}, while our knowledge at high redshift and luminosity remains limited. In fact, their detection requires high-quality X-ray spectra, which are challenging to obtain beyond the local Universe, with the notable exception of a few highly luminous and/or lensed QSOs \citep{Chartas03,Lanzuisi12,Vignali15,Dadina16,Bertola20}.
\cite{Chartas21} presented the first systematic characterization of UFOs at high redshift, based, however, on a small, highly biased sample of targets, mostly selected for being lensed (80\% of the sample) and/or with Broad/Narrow Absorption Lines in their UV spectra (80\% of the sample).

The WISSHFUL \xmm Multi-year Heritage Program (ID 094353, PI G. Lanzuisi) is one of the largest \xmm approved programs and aims to tackle this problem by obtaining deep \xmm observations, over three cycles (AO23-24-25), of a sample of 15 hyper-luminous ($\lbol > 10^{47}$ erg/s), high-redshift ($z\sim2-4$) quasars selected from the WISSH quasar sample \citep{Martocchia17,Degliagosti25}. These observations are specifically designed to reliably identify absorption features associated with nuclear winds/UFOs, investigate the wind-launching mechanism, and characterize the physical properties of nuclear winds in this critical luminosity and redshift regime.

\begin{table*}[t]
\begin{center}
\begin{small}
\caption{\xmm and \nustar observations breakdown.}
    \begin{tabular}{lccccccc}
\hline \noalign {\smallskip}
    Telescope & ObsID & Time & Tot. exp. & Instrument & Clean exp. & Net counts &  SNR \\ 
  & & & ks & & ks & 2-10 keV & 2-10 keV \\   
(1) & (2) & (3) & (4) & (5) & (6) & (7) & (8) \\  
    \hline \noalign {\smallskip}
    XMM0 & 0803950601 & 2017-11-17 & 33 &  pn      & 20 & 1568 & 36.8 \\ 
        &            &            &    &  MOS1/2   & 23 &  953 & 29.9 \\
    XMM1 & 0943530701 & 2024-10-12 & 65 &  pn      & 55 & 2341 & 42.8 \\ 
        &            &            &    &  MOS1/2   & 62 & 1534 & 35.6 \\  
    XMM2 & 0943530801 & 2024-10-22 & 30 &  pn      & 19 & 1118 & 31.4 \\ 
        &            &            &    &  MOS1/2   & 24 &  690 & 25.2 \\
    XMM3 & 0943532701 & 2024-10-31 & 24 &  pn      & 19 &  913 & 28.3 \\
        &            &            &    &  MOS1/2   & 23 &  572 & 23.0 \\  
    \hline \noalign {\smallskip}
  & & &  & & &  10-20 keV & 10-20 keV \\   
    \hline \noalign {\smallskip}
    NuSTAR1 & 61062007002 & 2024-10-21 & 74 & FPMA+B & 74 & 227 & 8.0 \\
    NuSTAR2 & 61062007004 & 2024-10-21 & 60 & FPMA+B & 60 & 225 & 8.9 \\
    \hline \noalign {\smallskip}
    \end{tabular}
    \vspace{-0.3cm}
    \label{tab:dataset_info}
\end{small}
\end{center}
\tablefoot{
(1) Obs. Name ; (2) Obs. ID; (3) Observation date; (4) Total exposure time; (5) Instrument; (6) Clean exposure time; (7) Net counts and (8) 
Signal-to-noise ratio in the $2-10$ keV band for \xmm and $10-20$ keV band for \nustar.}
\end{table*}

We also obtained hard X-ray coverage for the AO23 and AO24 WISSHFUL targets through simultaneous \nustar Cycle 10 (ID 61062, PI Lanzuisi) and Cycle 11 observations (ID 61162, PI Borrelli). This enables us to accurately constrain the high-energy continuum properties, such as the photon index, high-energy cutoff, and reflection fraction. These parameters are crucial not only for a more robust detection and characterization of UFOs, but also for understanding the coronal properties, such as optical depth and electron temperature, of these extreme AGN \citep{Dadina16,Kammoun17,Lanzuisi19,Kammoun23} and the potential connection with the presence of powerful outflows \citep{Tortosa24,Lanzuisi24}.
In fact, evidence is mounting that the presence and strength of such winds are somehow connected with the accretion properties of the central SMBH. An anti-correlation between BLR wind velocity and X-ray luminosity was found in the WISSH QSOs \citep{Zappacosta20,Degliagosti25}, while a similar connection with photon index was found in z>6 QSOs in the HYPERION survey \citep{Tortosa24}. Interestingly, faster large-scale ionized outflows seem to be also connected with soft X-ray spectra \citep{Wang16}. 

At the same time, a correlation between the photon index and Eddington ratio has been reported by several authors (\citealt{Shemmer08,Brightman13,Liu21} but see also \citealt{Trakhtenbrot17, Laurenti22, Tortosa22}). 
Recently, \cite{Inayoshi24} presented a theoretical framework for Comptonization in a super-Eddington accretion disk, powering nuclear radiation-driven outflows and resulting in optically thick, warm coronae that naturally lead to soft, weak X-ray emission. All these elements - high accretion rates, nuclear outflows, warm coronae, and soft emission - might be, in fact, related. 
Performing deep coordinated \xmm and \nustar observations of highly accreting luminous QSOs at high z is one of the best currently available approaches to solving this puzzle.

Here, we present the first results from our program focusing on the analysis of the \xmm and \nustar data for the first observed target, WISSH13 (RA=09:00:33.50, DEC=+42:15:47.00) at z=3.294 with $\lbol=8\times10^{47}$ erg/s \citep{Saccheo23}. The main properties of the target are summarized in \autoref{tab:one}.
The velocities of the BLR wind mapped by the CIV, and of the NLR wind mapped by the [OIII] emission line blue wings, are taken from \citealt{Vietri18} and \citealt{Bischetti17}, respectively. 
The target exhibits a powerful kpc-scale [OIII] outflow, 
and a CIV outflow with similar velocities
but a modest mass outflow rate, due to the smaller radius assumed for the BLR wind, and with large uncertainties due to the superimposition of the BLR and outflow components. For all these reasons, in the following, only the [OIII] outflow is used as a tracer of the galaxy-scale
outflow energetics.

WISSH13 is detected with LOFAR (150 MHz), uGMRT (670 MHz), and JVLA (3, 6, and 10 GHz) as unresolved, showing no hint of complex morphology on kpc scales (Fanelli et al. in prep.). The radio luminosity is among the highest in the WISSH sample ($L_{1.4 GHz} \sim 10^{25.5} W/Hz$), placing it nominally above the radio-loud threshold in absolute terms. However, using the classical definition of radio-loudness $\mathcal{R}={L_{5GHz}}/{L_{4400 {\AA}}}>10$ \citep{Kellermann89}, the source results radio quiet, with $\mathcal{R}=1.27$, due to the extreme optical luminosity.



The paper is organized as follows: \Cref{sec:reduction} describes the observations and data reduction; \Cref{sec:analysis} reports on the spectral analysis and UFO search carried out on the X-ray spectra; \Cref{sec:ufochar} describes the UFO characterization and energetics, while in \Cref{sec:discussion} we discuss the main results in the context of accretion/ejection models.
We adopt the cosmological parameters $H_0 = 70$ km s$^{-1}$ Mpc$^{-1}$, $\Omega_{\Lambda} = 0.73$ and $\Omega_m = 0.27$. 
Errors are given at a 90\% confidence level.

\section{Observations and Data Reduction}
\label{sec:reduction}

We obtained three \xmm observations for WISSH 13 in October 2024 (see \autoref{tab:dataset_info}). 

The data were reduced using SAS version 21, applying the optimized procedure of \citet{Piconcelli04} to maximize the signal-to-noise ratio (SNR) in a given band by iteratively selecting the optimal source extraction region and background threshold. This allowed us to retain $\sim80\%$ of the exposure time for Obs. XMM1, which was affected by moderate flaring during 70\% of the observation. Obs. XMM2 and XMM3 show more standard flaring levels. Final extraction radii are 15–30 arcsec for pn and 15–23 arcsec for MOS (70–85\% encircled energy at 1 keV). Ancillary response files were generated with arfgen, enabling applyabsfluxcorr=yes to improve the \xmm/\nustar cross-calibration \citep{Kang23}. MOS1 and MOS2 data are then combined into a single spectrum while the response matrices are averaged.

\begin{figure}[h]
    \centering
    \includegraphics[width=0.43\textwidth]{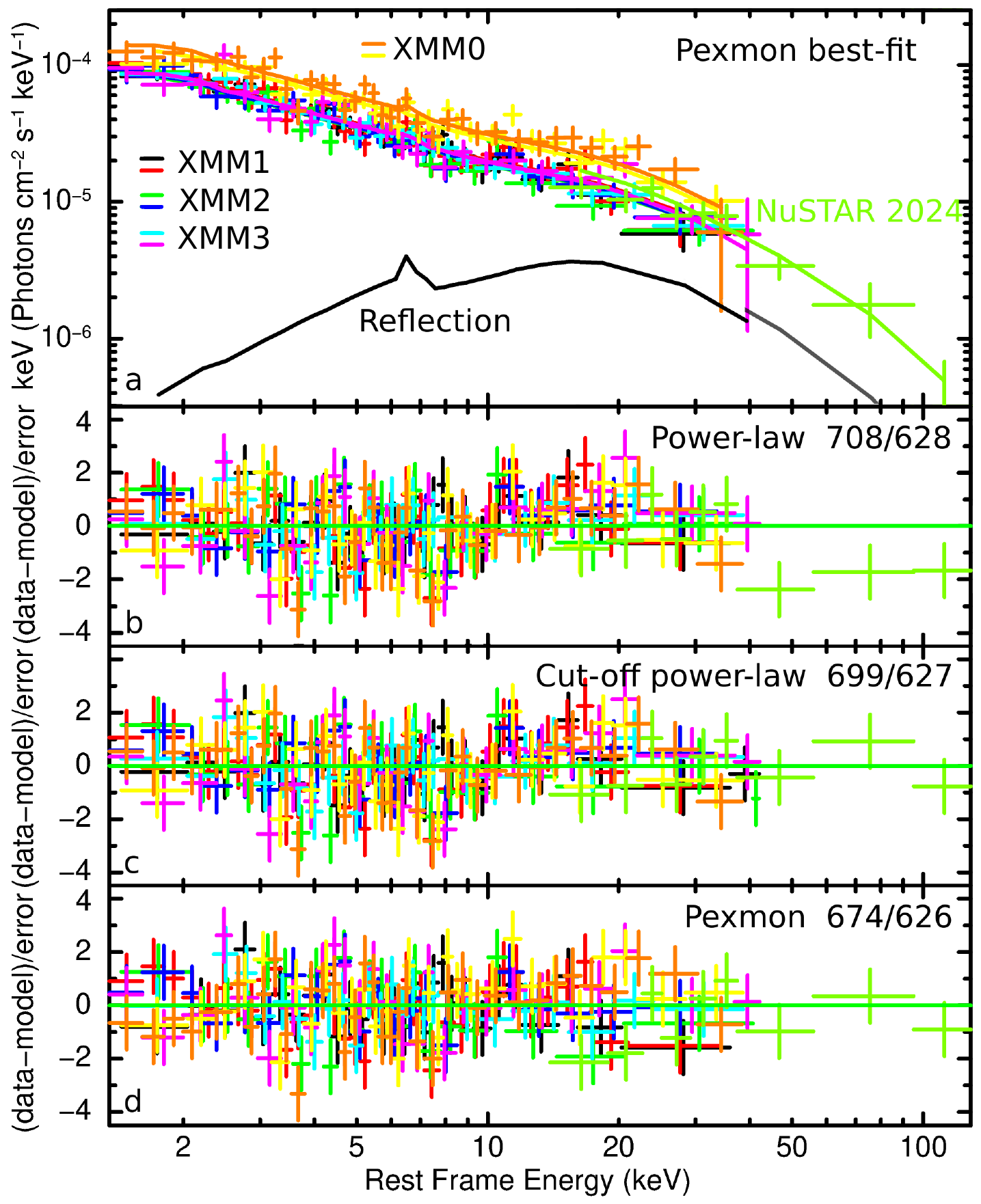}
    \caption{Panel $a$ shows the best-fit PEX model with the full \xmm and \nustar data set.
    pn and MOS1+2 data are shown with different colors for each observation, identified by its name.
    The reflection-only component is shown in black. Panels $b$, $c$, and $d$, show residuals for the PL, CPL and PEX models, respectively. Residuals for the REFL and CREFL models are not shown for clarity, as they are indistinguishable from those in panel d. C-stat$/$d.o.f. is reported for each model. \nustar FPMA and FPMB data are co-added for plotting purposes only. 
    }
    \label{fig:specdel}
\vspace{-0.3cm}
\end{figure}

We also included in the analysis Obs. XMM0 (PI: G. Risaliti), taken in 2017 (corresponding to a rest-frame separation of 1.6 years). The same reduction procedure was applied, yielding a net exposure of 20 ks. The source was $\sim1.5$ times brighter in 2017 than in 2024. Combined, the X-ray data presented here represent the highest quality X-ray spectra of a $z>2$ non-lensed QSO obtained so far, with $1.3\times10^4$ total 0.3-10 keV \xmm net counts and $5\times10^3$ in the 4-10 keV band rest-frame, comparable with local QSOs at z=0.1-0.5 in the SUBWAYS sample \citep{Matzeu23}.

As part of a \nustar Cycle 10 follow-up program, we obtained two simultaneous observations of 74 and 60 ks alongside XMM2. The data were processed with NuSTARDAS v2 (Heasoft v.6.30)\footnote{\url{https://heasarc.nasa.gov/lheasoft/}}) and CALDB v20220316 using standard filtering. Background spectra were extracted from an annular region (100–210 arcsec), while the source extraction radius of 50 arcsec ($\sim65\%$ encircled energy; \citealt{An14}) was chosen to optimize the SNR in the 3–20 keV band, yielding 520 total net counts and a SNR of 15.

\begin{table*}[t]
\begin{center}
\renewcommand{\arraystretch}{1.2} 
    \caption{Best-fit parameters for the continuum models.}
    \label{tab:fits}
    \begin{tabular}{lccccc}
        \hline
        Parameter              & PL & CPL & PEX & REFL & CREFL \\
        (1) & (2) & (3) & (4) & (5) & (6) \\
        \hline
        $\Gamma$               & $1.98_{-0.02}^{+0.02}$  & $2.00_{-0.03}^{+0.03}$ & $2.11_{-0.07}^{+0.08}$   & $2.10_{-0.03}^{+0.05}$ & - \\
        $E_\mathrm{cut}$ ~~ [keV] &  -     & $33_{-4}^{+9}$  & $63_{-18}^{+38}$  & $79.5_{-15.3}^{+28.9}$ & - \\
        $R$                    &  -     & -   & $1.8_{-0.7}^{+0.9}$ & $1.4_{-0.6}^{+0.7}$      & $1.7\pm0.5^a$ \\
        $kT_e$ ~~~ [keV]           &  -     & -   &  -    &  -      &  $17.4_{-4.3}^{+17.7}$ \\
        $\tau$                 &  -     & -   &  -    &  -       & $1.4_{-0.7}^{+0.4}$ \\
        C-stat$/$d.o.f.        & $708/628$ & $699/627$ &  $674/626$ & $681/626$ & $679/626$ \\
        $p_{\rm MC}~^b$ &  0.02 & 0.31 & 0.44 & 0.35 & 0.45 \\
        Model sign.$^c$    & - & 99.7 & $>99.9$ & $>99.9$ & $>99.9$\\
        \hline
        $F_{0.5-10}~^d$ [erg/s/cm$^2$] & \multicolumn{5}{c}{$2.07\pm0.09~(3.5\pm0.2)\times10^{-13}$} \\
        $L_{2-10}~^d$ ~~~[erg/s]   &  \multicolumn{5}{c}{$1.11\pm0.03~(1.72\pm0.06)\times10^{46}$}  \\
        \hline 
        \vspace{-0.7cm}
\end{tabular}
\end{center}
\tablefoot{
(1) Model parameters referring to: (2) the power-law model; (3) the cut-off power-law model; (4) the \texttt{pexmon} model; (5) the \texttt{xillver} model, and (6) the \texttt{compTT+xillverCP} model.
$^a$ Reflection fraction computed from the flux ratio between the primary and reflected components in the 20-40 keV rest frame band and scaled to R units.
$^b$ Null Hypothesis Probability from Monte Carlo simulations. The simple power law model is rejected at 98\% confidence level, while the other models are all statistically acceptable.
$^c$ Significance in \% of the improvement over the simpler model. See text for details.
$^d$ Average 2024 (2017) 0.5-10 keV flux and 2-10 keV rest-frame luminosity. They are consistent across the different models.}
\vspace{-0.3cm}
\end{table*}

The \xmm and \nustar spectra were binned following the \cite{KaastraBleeker16} binning scheme to ensure optimal spectral resolution given the source and background flux levels (see \citealt{Matzeu23,Lanzuisi24}). The spectra were analyzed using the \xspec software package (version 12.14.1 \citealt{Arnaud96}). The $\cstat$ statistics \citep{Cash79}, with background subtraction (W-stat in \xspec) was adopted during spectral fitting.

\section{Spectral Analysis}
\label{sec:analysis}

After verifying that the continuum spectral
shape is consistent between the two epochs, we performed a joint broadband spectral analysis of WISSH13, simultaneously fitting the 2024 and 2017 \xmm pn and MOS data, together with the 2024 \nustar FPMA and FPMB data. 
In all the models described below, we included a multiplicative constant component, free to vary across data sets, to account for cross-calibration and variability between the \xmm and \nustar cameras. The value of this normalization constant was always between 0.9 and 1.1 between \xmm pn and MOS and \nustar FPMA and FPMB cameras in the 2024 observations, in line with expected cross-calibration systematic uncertainties, while the 2017 \xmm data have systematically higher normalization due to long-term flux variation. We also include Galactic absorption, modeled with a fixed $\nhgal=2.3\times10^{20} \cmsq$ \citep{HI4PI16}, while intrinsic absorption was not needed in all models. 

\subsection{Continuum Modeling}
\label{sec:continuum}

\begin{figure*}[t]
    \centering
    \hspace{-0.2cm}\includegraphics[width=0.4\textwidth]{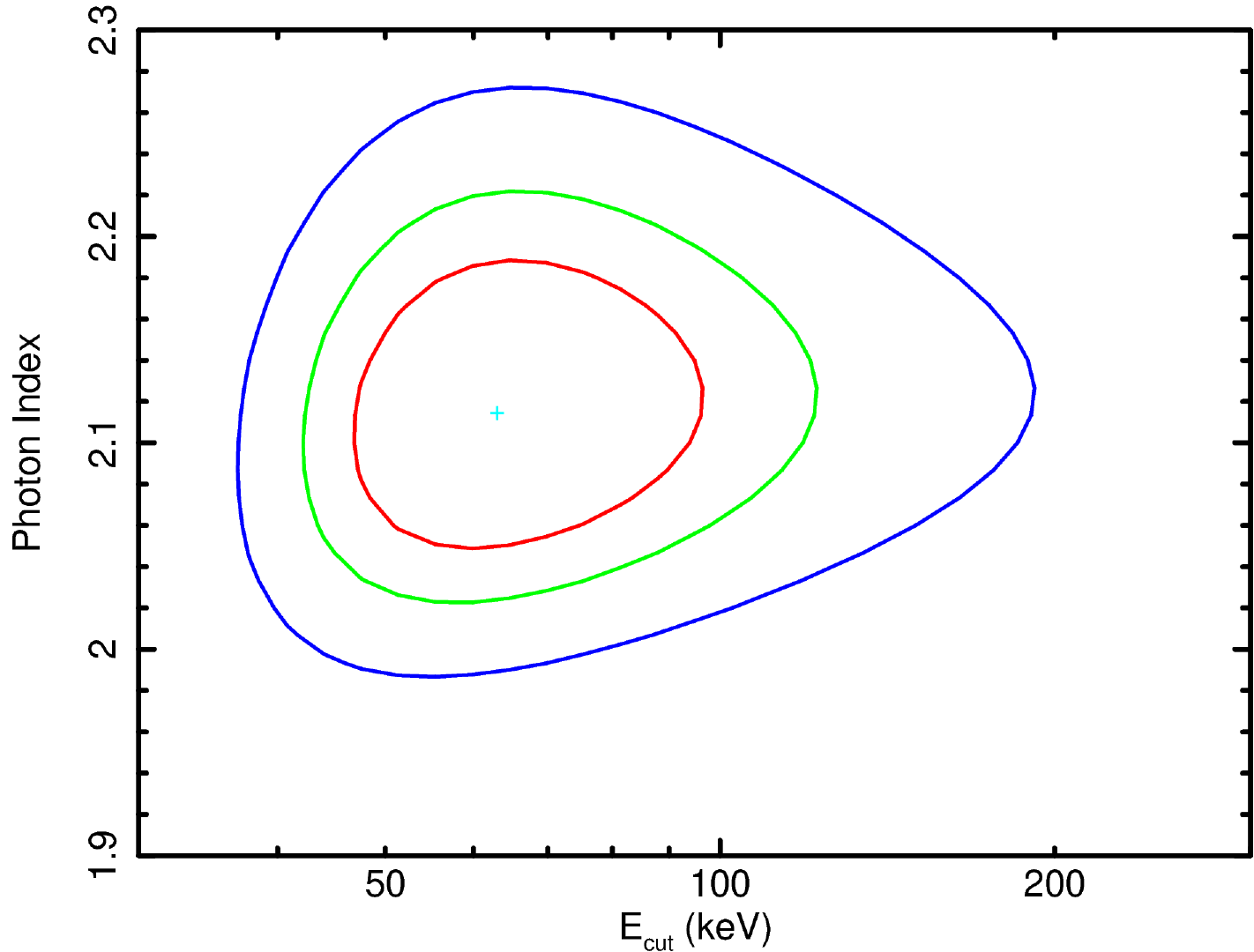}
    \hspace{0.9cm}
    \includegraphics[width=0.4\textwidth]{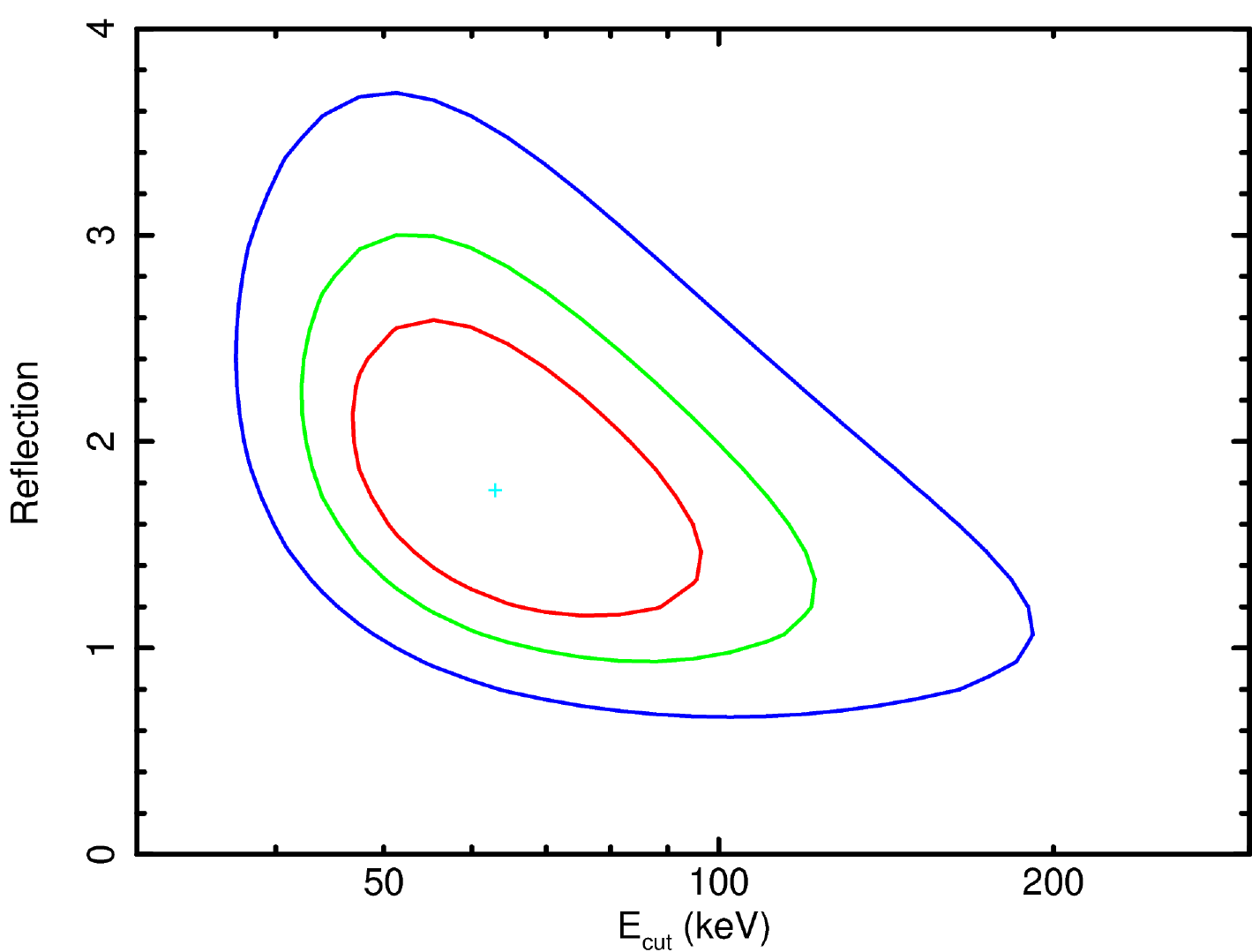}\vspace{0.5cm}
    \includegraphics[width=0.4\textwidth]{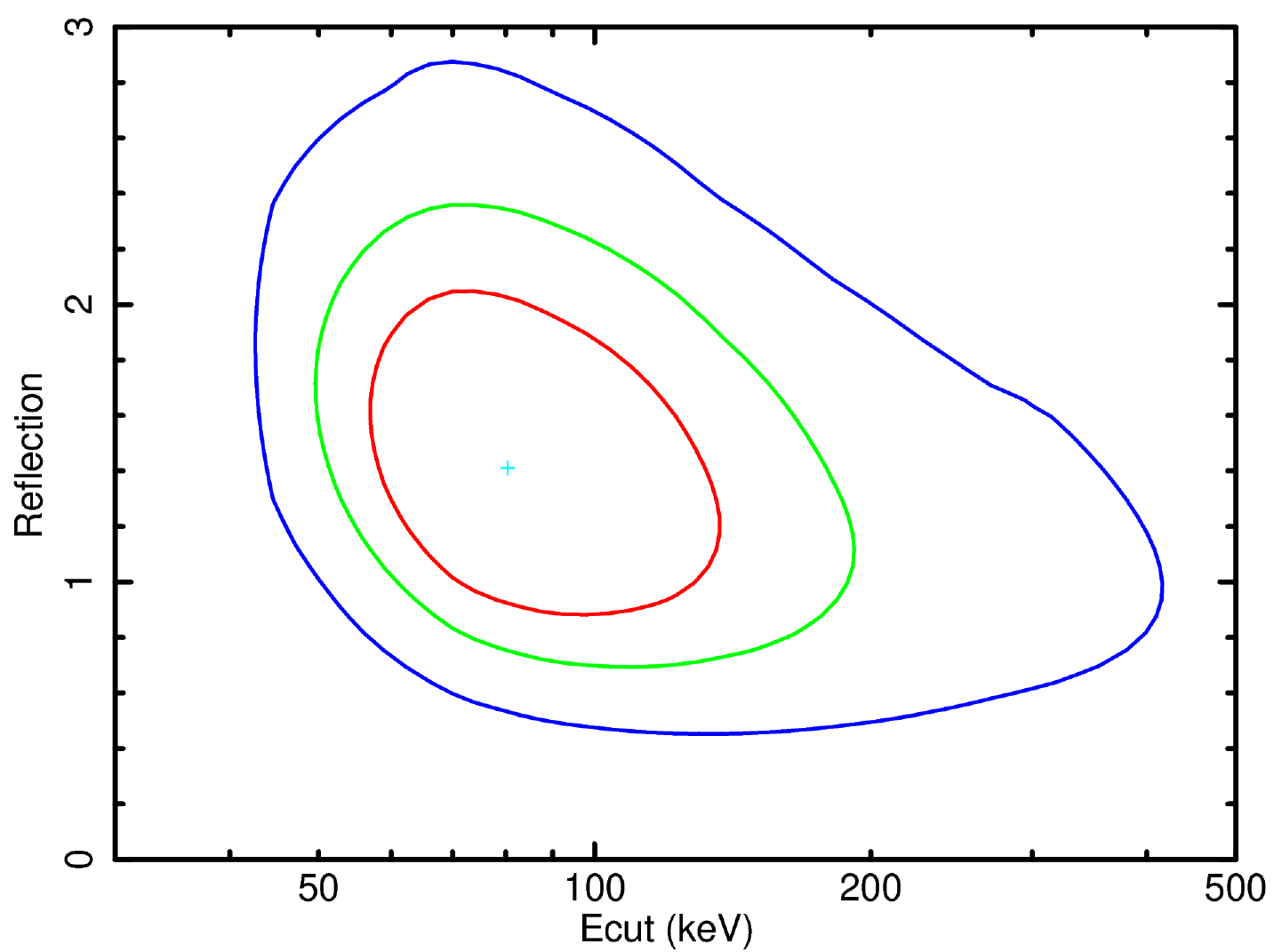}
    \hspace{0.75cm}
    \includegraphics[width=0.4\textwidth]{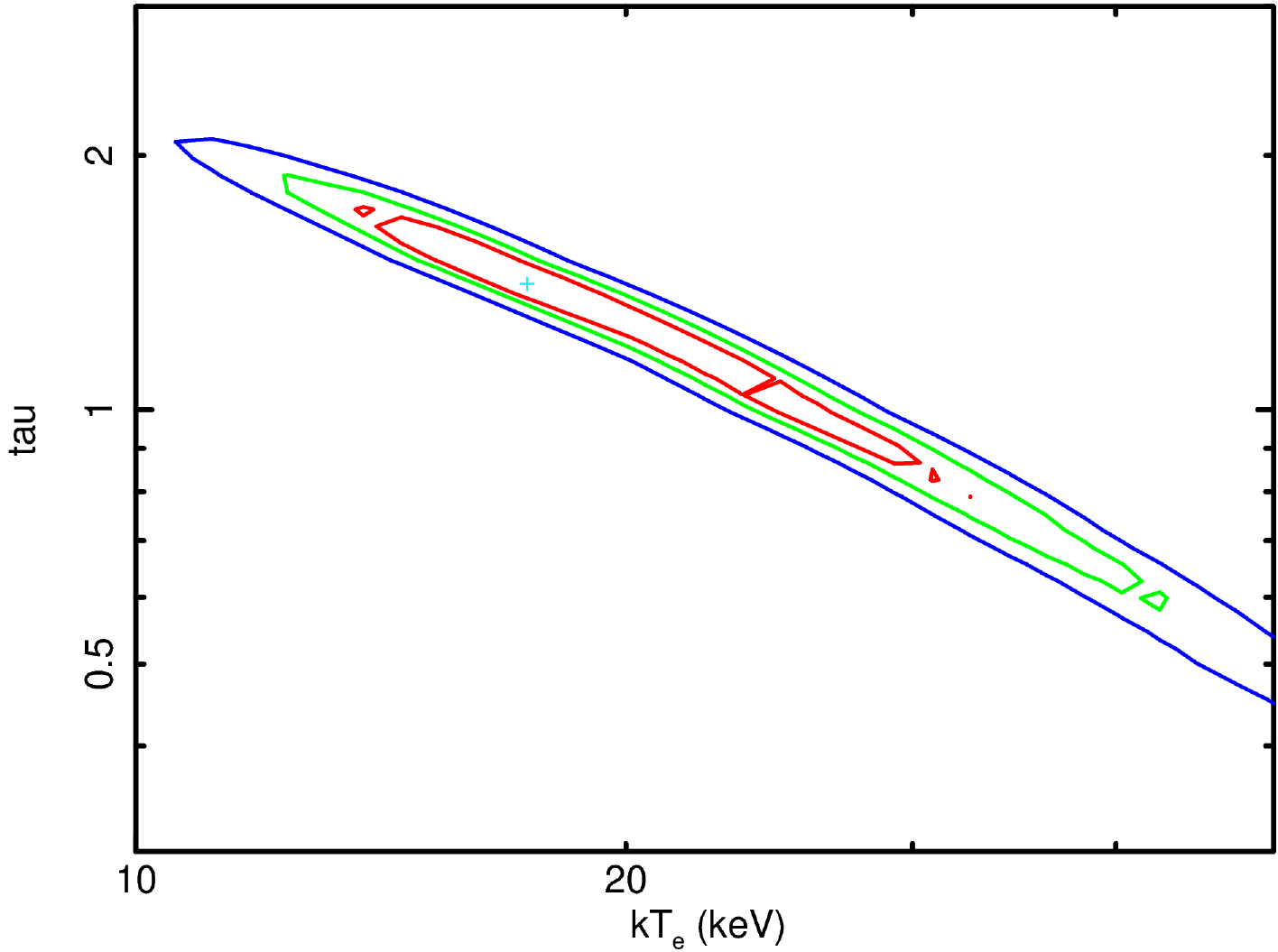}    
    \caption{Top row: Confidence contours for the PEX model, $E_\mathrm{cut}$ vs $\Gamma$ (left) and $E_\mathrm{cut}$ vs $R$ (right). 
    Bottom row: Confidence contours for $E_\mathrm{cut}$ and $R$ for the REFL model (left) and for $kT_e$ and $\tau$ for the CREFL model (right). The latter reveals a strong $\tau-kT_e$ degeneracy: higher $kT_e$ is compensated by lower $\tau$ (and vice versa), to reproduce the same observed photon index $\Gamma$.
    Red, green, and blue contours show 68, 90, and 99\% confidence levels, respectively.}
    \label{fig:cont_contours}
    \vspace{-0.3cm}
\end{figure*}

We fitted the combined \xmm and \nustar spectra using several physically motivated models, listed below. The adopted continuum models are:
i) a simple power-law model (\texttt{tbabs$*$zpowerlw}, PL) with free parameters the photon index $\Gamma$ and normalization; 
ii) a power law with high-energy cut-off at the source redshift, cut-off power-law (\texttt{tbabs$*$cutoffpl}, CPL), with free parameters $\Gamma$, the high-energy cut-off $\ecut$, and normalization;
iii) a cold reflection model using pexmon \citep{Nandra07}, (\texttt{tbabs$*$pexmon}, PEX) with free parameters $\Gamma$, $\ecut$, the reflection parameter R ($>0$ so that the model also includes the primary power-law continuum) and normalization; 
iv) a cold reflection model using xillver \citep{Garcia10,garcia13}, (\texttt{tbabs$*$xillver}, REFL) with free parameters $\Gamma$, $\ecut$, R ($>0$ as for PEX, and normalization, where the ionization parameter is set to $log(\xi)=0$, given the lack of evidence for ionized reflection; 
v) a Comptonization plus reflection model using compTT  \citep{Titarchuk94} (\texttt{tbabs$*$(compTT + xillverCP)}, CREFL) with free parameters the coronal electron temperature kT$_e$, and the coronal optical depth $\tau$. The geometry is set to 'slab' and the disk seed photon temperature to 100 eV. To reproduce the reflection in this model, we added a xillverCP component in reflection ($R=-1$), where the electron temperature of the corona is linked to the temperature in compTT, the normalization is free to vary, the photon index is set to $\Gamma=2.0$, consistent with the observed value from previous models, and the ionization parameter $log(\xi)=0$. 
All the reflection components have the inclination angle parameter set to $30\degree$, an average value for face-on, type-1 QSOs.

For the PEX, REFL and CREFL models, the Fe abundance parameter, which regulates the Fe emission lines strength with respect to a given reflection spectrum, is set to 0.5 Solar. This choice was motivated by the weak emission observed at 6.4 keV: the equivalent width of a narrow Gaussian line ($\sigma=10$ eV) added to the simple PL model and centered at 6.4 keV is $<60$ eV rest-frame. Indeed, when left free to vary, the Fe abundance best-fit gives an upper limit of A$_{\rm Fe}\lesssim0.6$ Solar in all models. 

Figure~\ref{fig:specdel} compares the spectral residuals of the various models tested, showing that the three models, including both high-energy cut-off and reflection, all provide a good description of the data. We report the best-fit parameters and statistics of each model in Table~\ref{tab:fits}. 
We evaluated the goodness-of-fit using Monte Carlo simulations (1000 realizations) with the Anderson-Darling test statistic using the \texttt{goodness} command in \xspec. The simple power-law model is rejected at the 98\% confidence level ($p_{\rm MC}=0.02$).
The inclusion of a high-energy cut-off significantly improves the fit, yielding a model that is formally acceptable ($p_{\rm MC}=0.31$ corresponding to 69\% of realizations with lower observed statistic).

To quantify the significance of the additional continuum components, we employed the Likelihood Ratio Test (LRT), assuming that the $\Delta \cstat$ approximately follows a $\chi^2$ distribution with degrees of freedom equal to the number of added parameters \citep{Cash79}.
The addition of the high-energy cut-off going from PL to CPL, reduces the C-statistic by $\Delta \cstat=9$. This improvement is significant at the $\>99.7\%$ level for the addition of one parameter.
The further addition of a reflection component, going from CPL to either with PEX, REFL, or CREFL model, reduces the C-statistic by $\Delta \cstat=18-25$. 
In this case, the improvement is significant at the $\>99.9\%$ level for one parameter, for all three models. Thus, while the cutoff power-law provides a plausible global description of the continuum, the reflection component is statistically required to explain the spectral curvature in the hard X-ray band. 
We note that the lower $\ecut$ value obtained with the simple cutoff power-law model ($\sim33$ keV) compared to the reflection models ($\sim60-80$ keV) is expected, as the spectral curvature introduced by the unmodeled reflection component is translated into an artificially lower cut-off energy \citep[see e.g.,][]{Tortosa18a,Middei19}.

We also investigated parameter degeneracies using contour plots. Figure~\ref{fig:cont_contours} shows the confidence contours for $\Gamma$ vs. $\ecut$ and R vs. $\ecut$ for the PEX model in the top row. 
The three parameters are all well constrained, with the photon index being $\gtrsim2$, $R\gtrsim1$, and $\ecut\lesssim200$ keV at $99\%$ confidence level.
The bottom row displays the R vs. $\ecut$ contours for the REFL model (left). This model shows similar reflection and cut-off to PEX. Finally, the bottom right panel shows $\tau$ vs. kT$_e$ for the CREFL model. As expected for thermal Comptonization, these two parameters are tightly anti-correlated, along a curve of nearly constant Compton {\it y} parameter ($\propto kT_e\times\tau$) that reproduces the observed spectral slope $\Gamma$ \citep{Tortosa18a,Middei19}.
\begin{figure}[t]
    \centering
    \includegraphics[width=0.4\textwidth]{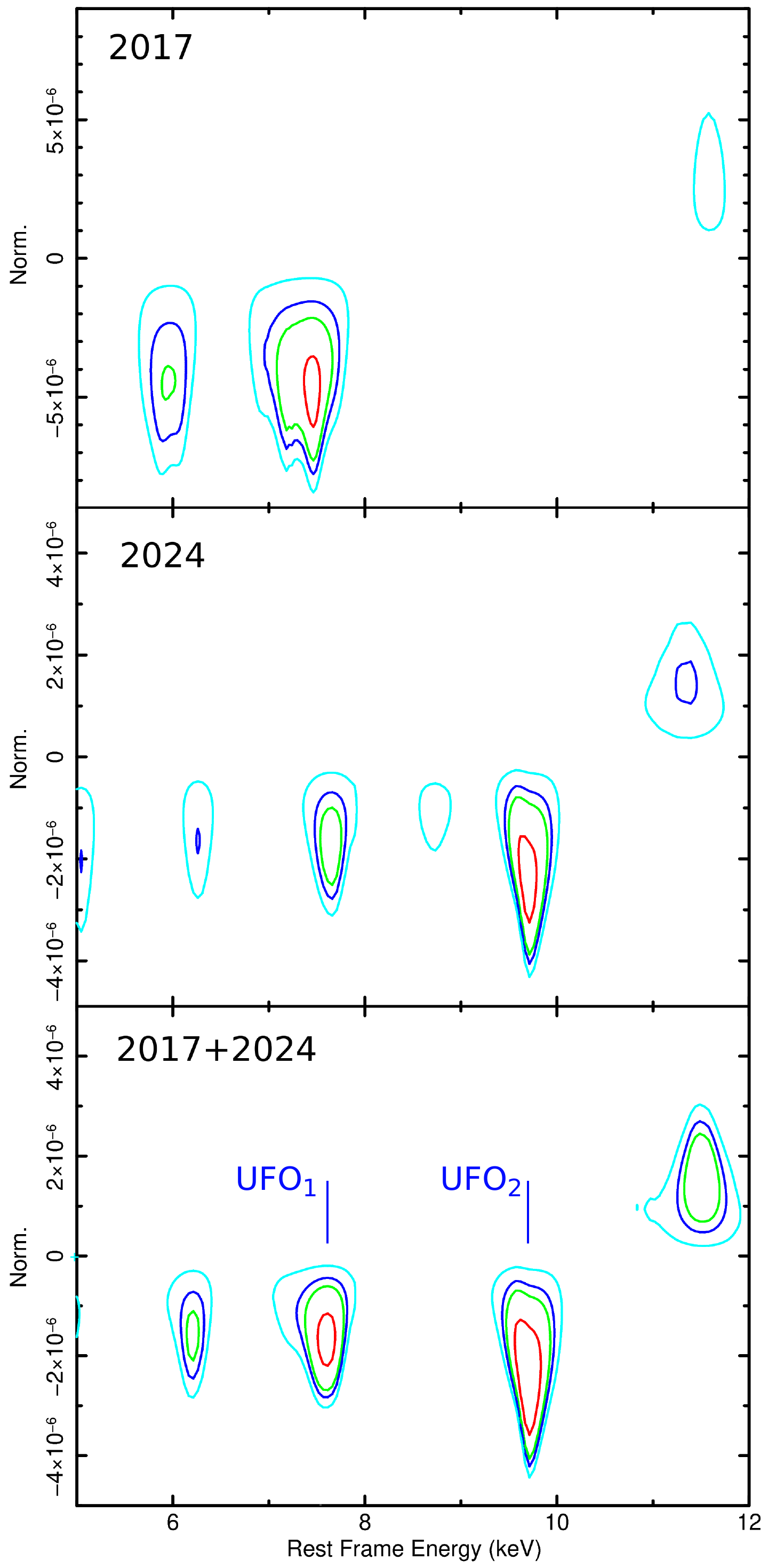}
    \caption{Blind line scans over the 5--12 keV rest frame band. The plot shows contours of $\Delta C$ in the energy vs. line-normalization parameter space for a narrow Gaussian ($\sigma=10$ eV) added to the baseline continuum. Cyan, blue, green, and red contours correspond to a $\Delta C$ of -2.3, -4.61, -5.99, and -9.21, respectively, which can be translated into a significance of 68, 90, 95, and 99\% for two parameters.}
    \label{fig:line_scan}
    \vspace{-0.3cm}
\end{figure}

The high-energy cut-off from PEX and REFL ($\sim60-80$~keV) and the electron temperature from CREFL ($\sim20$~keV) agree within the errors with the typical scaling factor of $2-3$ \citep{Petrucci01,Middei19} and consistently show that WISSH13 has one of the coolest coronal temperatures ever measured \citep{Kara17,Tortosa17,Reeves21,Bertola22,Lanzuisi24}.
The source accretion rate close to or above Eddington ($\eddratio\sim3$, see \autoref{tab:one}) is in line with the observed trend of cooler coronae for highly accreting SMBHs observed in \citet{Peluso26sub} (but see also \citealt{Zhao2026}) on a sample of Seyferts and QSOs from the local Universe to Cosmic Noon,  covering a wide range of BH masses ($log(\mbh/[\Msun])=6-10$) and luminosities ($log(\lbol/[\rm erg\,s^{-1}])=42-48$).

\subsection{Blind line search}
\label{sec:linescan}

We searched for the presence of any UFO-related absorption features superimposed on the best-fit continuum models described above. To estimate the energy, width, and significance of any absorption or emission feature relative to the underlying continuum, we conducted a blind search in the simultaneous fit of \xmm and \nustar spectra, i.e., a blind Gaussian line scan across the Fe K band. We refer to \citet{Miniutti06}, \citet{Tombesi10}, \citet{Gofford13}, \citet{Matzeu23} and \citet{Lanzuisi24} for a similar approach. 

\begin{figure}[t]
\centering
\includegraphics[width=0.85\linewidth]{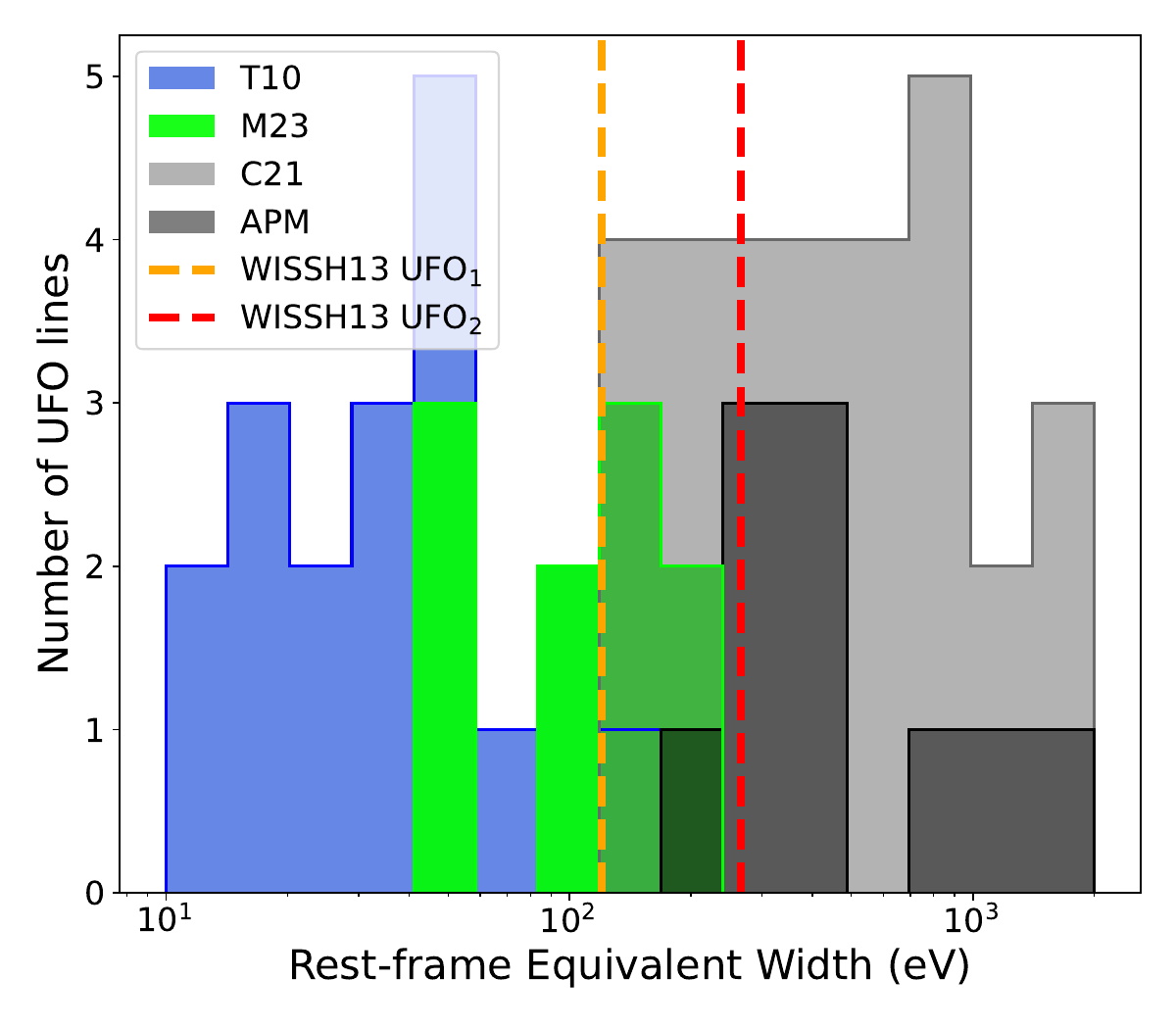}
\caption{Rest-frame equivalent widths for WISSH13 (orange and red dashed lines for UFO$_1$ and UFO$_2$, respectively) compared with Seyferts at $z<0.1$ in blue (\citealt{Tombesi10}, T10), QSOs at $z=0.1-0.5$ in green (\citealt{Matzeu23}, M23) and lensed, high-$z$ QSOs in gray (\citealt{Chartas21}, C21, and references therein),
including the multiple detections in APM~08279+5255 \citep{Chartas09}, shown in black.}
\label{fig:EW_comparison}
\end{figure}

We added a narrow Gaussian line (with line width fixed
to $\sigma = 10$ eV, i.e., unresolved) to the underlying continuum modeled with PEX, which has the lowest $\cstat$ value. However, we verified that consistent results are obtained by performing the same scan on the REFL or CREFL, i.e., the observed residuals are independent of the details of the primary emission and reflection spectral shapes.
We allowed the power-law photon index and normalization to vary during the search. The reflection parameter and high-energy cut-off were fixed to their best-fit values to reduce computational time. 
The scan was performed over the rest–frame $5-12$ keV interval for both emission and absorption features.

Since absorption features due to UFOs are known to be highly variable in time, changing centroid energy (i.e., velocity), and appearing/disappearing both on short \citep[e.g.,][]{Cappi06,Chartas18IRAS13224,Braito21} and long time scales \citep[e.g.,][]{Saez11,Matzeu17,Bertola22}, we first performed the line scan on the \xmm 2017 and 2024 spectra separately. In both cases, the 2024 \nustar spectra are included to help constrain the overall continuum shape and reflection intensity, and the best fit from the merged 2017+2024 spectra is used as a baseline. The resulting $\Delta C$ improvements over the baseline fit as a function of energy are shown in \autoref{fig:line_scan} for the 2017 and 2024 spectra in the top and middle panels.

Two significant absorption features emerge: one at $\sim7.5$ keV in the rest frame ($\sim1.8$~keV observed frame), detected at 99\% and 95\% confidence in 2017 and 2024, respectively, and one at $\sim9.7$ keV ($\sim2.3$~keV observed frame), seen only in the 2024 spectra. 
For the $\sim7.5$ keV feature, we performed a merged 2017+2024 scan to fully exploit the constraining power of the entire dataset, while for the $\sim9.7$ keV feature, we ignored the 2017 spectrum between 9 and 10.5 keV.
The results are shown in \autoref{fig:line_scan}, bottom panel. By fitting a narrow, unresolved Gaussian line ($\sigma=10$ eV) to the two strongest features in the joint fit, we obtain best fit values for the first one (hereafter UFO$_1$) $E_{\rm rest}=7.6_{-0.2}^{+0.1}$ keV with $\Delta C=9.8$ and a single-trial F–test probability of $98.97\%$. The observed-frame equivalent width is $\mathrm{EW}=28_{-11}^{+15}$ eV, corresponding to $\sim120$ eV in the rest frame. The second feature (UFO$_2$ hereafter) lies at $E_{\rm rest}=9.8_{-0.2}^{+0.1}$ keV with $\Delta C=11.7$ and an F–test probability of $99.25\%$; its observed-frame equivalent width is $\mathrm{EW}=62_{-29}^{+35}$ eV (about $260$ eV rest–frame). 
The best-fit values for the two Gaussian components are reported in \autoref{tab:ufo_gaussians}, where we also report the constraints on the line width obtained by leaving $\sigma$ as a free parameter. 

\begin{table}[t]
\centering
\caption{Best-fit parameters of the Gaussian components for UFO$_1$ and UFO$_2$ and statistical significance of the features.}
\label{tab:ufo_gaussians}
\renewcommand{\arraystretch}{1.35}
\begin{tabular}{lcc}
\hline\hline
 & \textbf{UFO$_1$} & \textbf{UFO$_2$} \\
\hline
$E$ (keV)            & $7.6_{-0.2}^{+0.1}$ & $9.8_{-0.2}^{+0.1}$ \\
$\sigma$ (keV)        & $<0.39$ ($<1.67$)    & $<0.29$ ($<1.24$)\\
Intensity                & $-1.7_{-0.9}^{+0.7}\times10^{-6}$ & $-2.1_{-1.2}^{+1.0}\times10^{-6}$ \\
EW (eV)              & $28_{-11}^{+15}$ (120)   & $62_{-29}^{+35}$ (266) \\
$\Delta C$-stat      & 9.8        & 11.7 \\
$P_{\rm F\text{-}test}$ & 99.0\%     & 99.3\% \\
$P_{\rm MonteCarlo}$ & 96.7\%  & 98.9\% \\
\hline
\end{tabular}
\tablefoot{
Line energy is rest-frame; line width $\sigma$ and equivalent width EW are in the observed frame, with rest-frame values in parentheses. $\sigma$ upper limits at 90\% confidence level.}
\end{table}

\begin{figure*}[t]
    \centering
    \includegraphics[width=0.40\linewidth]{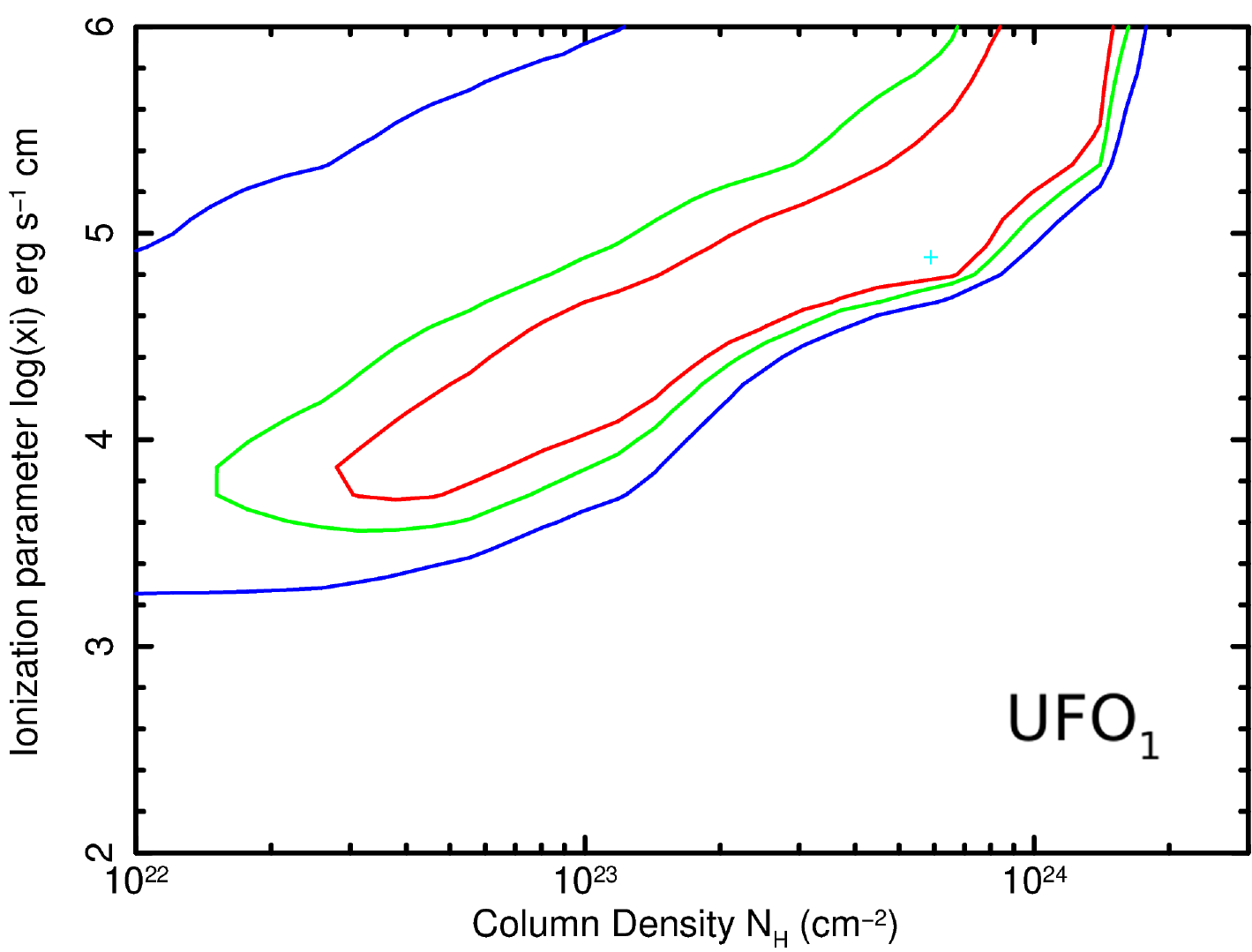}
    \hspace{0.8cm}
    \includegraphics[width=0.40\linewidth]{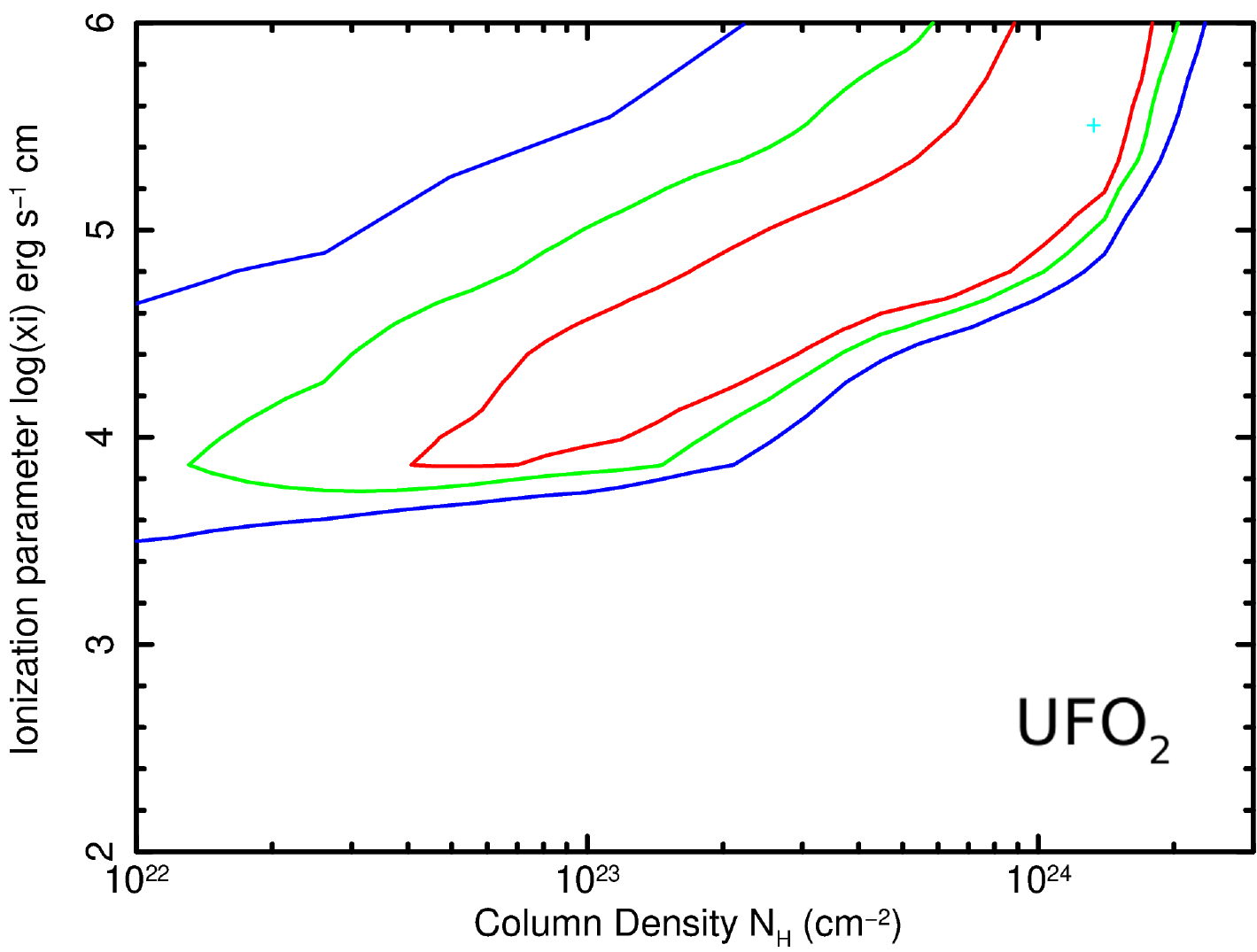}
    \caption{Confidence contours (at 68, 90, 99\% confidence levels) for $\log N_{\rm H}$ vs. $\log\xi$ for UFO$_1$ (left) and UFO$_2$ (right) from the fit with \textsc{xstar} tables with $v_{\rm turb}=5000$ km s$^{-1}$.}
    \label{fig:contours}
\end{figure*}

\begin{figure*}[h]
    \centering
    \includegraphics[width=0.40\linewidth]{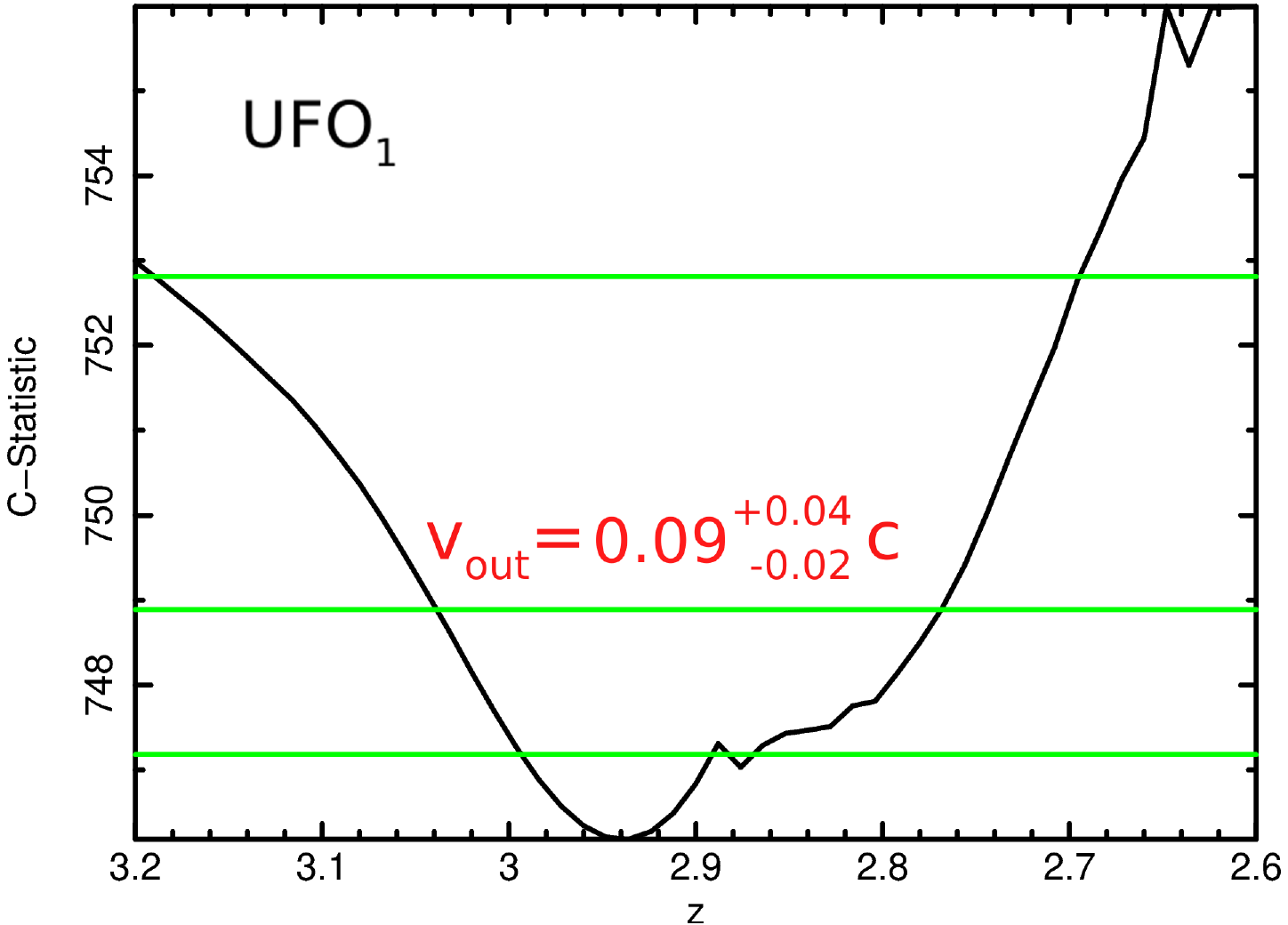}
    \hspace{0.8cm}
    \includegraphics[width=0.40\linewidth]{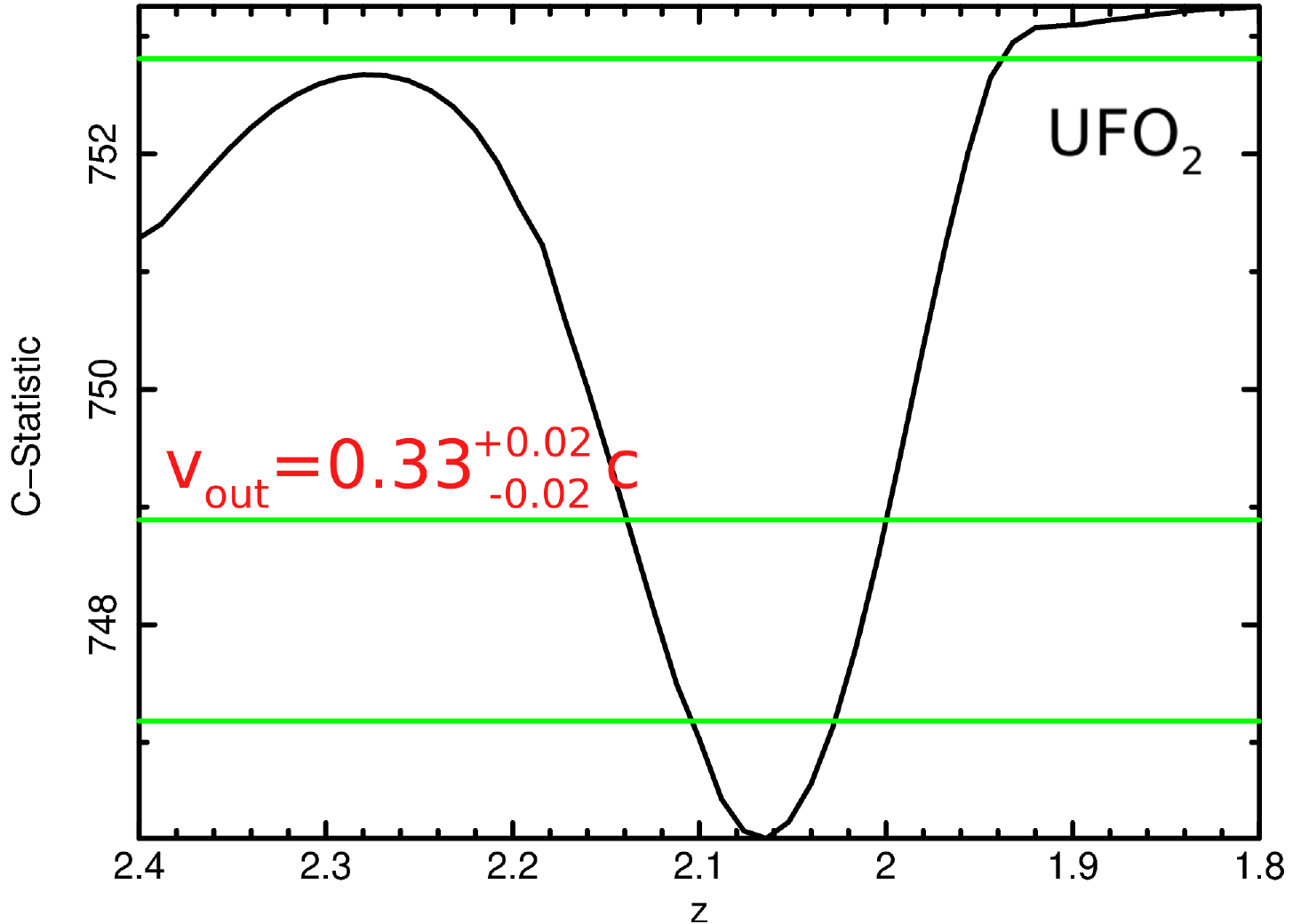}
    \caption{Redshift contours from the fit with \textsc{xstar} tables with $v_{\rm turb}=5000,{\rm km,s^{-1}}$, for UFO$_1$ (left) and UFO$_2$ (right). The horizontal green lines mark the 68, 90, and 99\% confidence levels. The 90\% constraints on $z$ are converted to $v_{\rm out}$ values quoted in the red labels (see text for details).}
    \label{fig:zcontours}
    \vspace{-0.3cm}
\end{figure*}

Single-trial F–test probability is known to overestimate the significance of line detections, as it does not consider the
full energy range over which a line might be observed by chance,
the so-called {\it look-elsewhere} effect \citep{Protassov02}. Therefore, a full set of Monte Carlo (MC) simulations was performed to properly assess the significance of the detected features.
Details are reported in \autoref{app:mc}. Briefly, we simulate line-free \xmm and \nustar spectra from the best-fit continuum with the same set of exposures of the real data, re-fit them with and without a narrow Gaussian line free to vary in intensity (both positive and negative) and energy in the full $5-12$ keV range, and store the best-fit $\cstat$ values with and without it. This shows the distribution of spurious $\Delta C$ peaks over the 5–12 keV range. Comparing the observed $\Delta C$ values to this distribution yields global confidences of $96.73\%$ for UFO$_1$ and $98.86\%$ for UFO$_2$, 
and therefore both features can be considered statistically significant, following the threshold adopted in e.g., \citet{Tombesi10,Gofford13,Matzeu23}.

We verified that the detected absorption features are not driven by instrument systematics in the observed $\sim1.8-2.3$ keV range. In particular, fitting the pn and MOS spectra separately yields consistent line energies and comparable improvements in the fit statistic, and
the result remains stable against variations in the source/background extraction choices and flare-screening thresholds. In addition, no analogous residual is present in the corresponding background spectra around the line energies.

\autoref{fig:EW_comparison} compares the observed EWs of UFO$_1$ and UFO$_2$ with literature samples. The WISSH13 features lie
at the upper limit of the distributions for low-$z$ Seyferts ($z<0.1$, \citealt{Tombesi10}) and local QSOs ($z=0.1-0.5$, \citealt{Matzeu23}), and below the larger EWs commonly reported at high redshift ($z=1-4$, \citealt{Chartas21} and references therein). 
This highlights the importance and power of long, dedicated spectroscopic campaigns in high-z QSOs to go beyond the detection of the most extreme UFOs we have known so far, and explore what is most probably the bulk of the UFO population, showing properties analogous to the low-z AGN population, or alternatively test whether there is a real evolution in redshift and/or luminosity of such properties. 
We note that 1/3 of the EW measurements for the high-z sample in \citet{Chartas21} come from the persistent pair of UFO features observed in the extreme case of APM~08279+5255 \citep{Chartas09}, which may further bias the distribution.

\section{Properties of the UFO}

\begin{figure*}[t]
    \centering
    \includegraphics[width=0.4\linewidth]{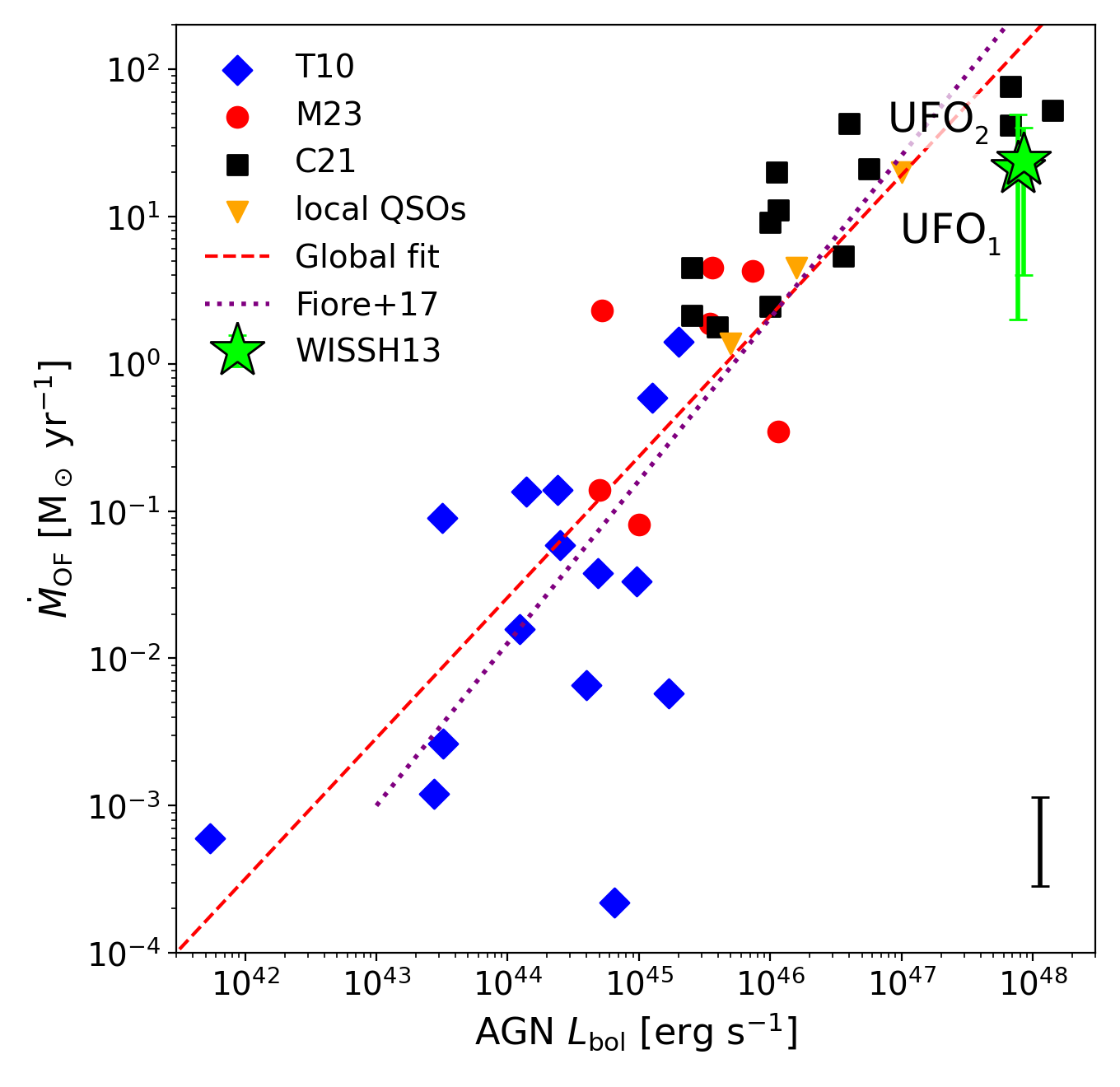}
    \hspace{0.5cm}
    \includegraphics[width=0.4\linewidth]{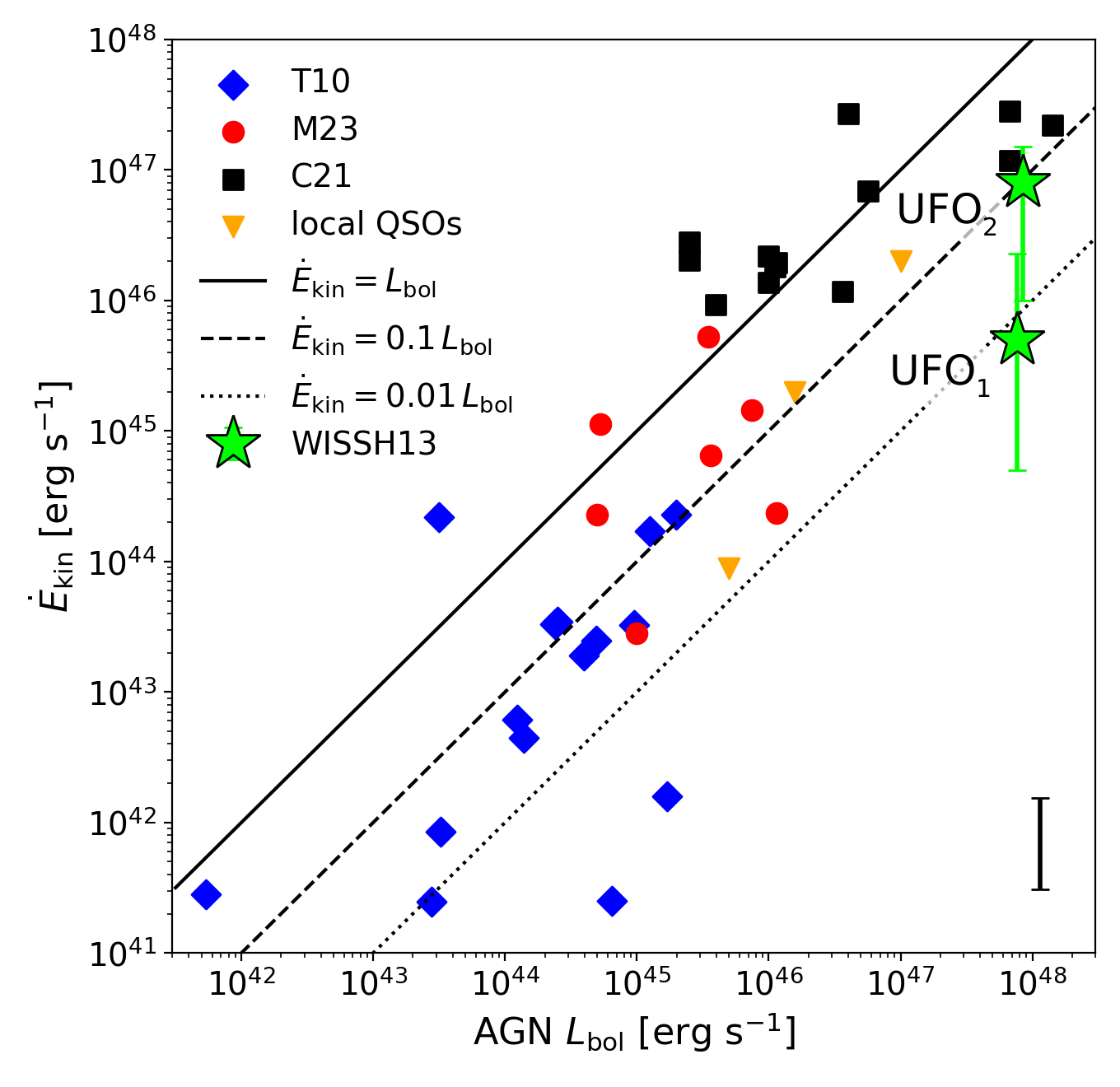}
    \caption{UFO energetics for WISSH13 and literature samples: local Seyferts (T10, blue diamonds), low-z QSOs from SUBWAYS sample (M23, red circles), high-z lensed QSOs (C21, black squares) and local QSOs (PDS~456, IRAS~11119 and Mrk~231, see text for references, orange triangles).
    Left: $\mout$ vs. $\lbol$. The dotted purple line is the relation derived in \cite{Fiore17}, while the red dashed line shows the linear fits obtained by jointly fitting all our data points.
    Right: $\ekin$ vs. $\lbol$. Lines for $\ekin=1,0.1,0.01\lbol$ are shown as continuous, dashed, and dotted lines, respectively. The average error bar on $\mout$ and $\ekin$ is shown on the bottom right of each plot.}
    \label{fig:energetics}
    \vspace{-0.3cm}
\end{figure*}

We model the two absorption complexes with photoionized absorption tables computed with \xstar, adopting a power-law ionizing SED matched to the measured continuum, with $\Gamma=2$, in line with the observed value, and $\lion=6\times10^{46}$ erg~s$^{-1}$,  extrapolated from the observed spectrum to the 1-1000 Ryd band, i.e., $0.0136 - 13.6$ keV. We explored the parameter space $\nh=[10^{20}-3\times10^{24}] \cmsq$ in 20 steps and $\logxi=[1-6]$ in 20 steps.
We used as input parameters a gas density of $10^9$ cm$^{-3}$, a temperature of $1\times10^6$ K, and set the remaining parameters to their default values within \xstar.
After several tests we adopted a table with input turbulent velocity $v_{\rm turb}=5000$ km s$^{-1}$, which is consistent with what is obtained in a fit with turbulent velocity as a free parameter ($\sim4800$ km/s for UFO$_1$ and $\sim4500$ km/s for UFO$_2$) and is in agreement with the upper limits on $\sigma$ shown in \Cref{tab:ufo_gaussians}: $v_{\rm turb}=5000$ km s$^{-1}$ corresponds to a $\sigma$ of $\sim40-50$ eV observed-frame at 2~keV.

\label{sec:ufochar}
\begin{table}[t]
\centering
\caption{\xstar table best-fit parameters. }
\label{tab:xstar_params}
\renewcommand{\arraystretch}{1.20}
\begin{tabular}{lcc}
\hline\hline
 & \textbf{UFO\_1} & \textbf{UFO\_2} \\
\hline
$\log N_{\rm H}$ (cm$^{-2}$) & $23.8_{-1.4}^{+0.5}$ & $24.1_{-0.9}^{+0.3}$ \\
$\log \xi$ (erg cm s$^{-1}$) & $>3.7$ & $>3.8$ \\
$z_o$                            & $2.94_{-0.17}^{+0.09}$ & $2.06_{-0.07}^{+0.07}$ \\
$v_{\rm out}$ ($c$)            & $0.09^{+0.04}_{-0.02}$ & $0.33^{+0.02}_{-0.02}$ \\
$\Delta C$-stat & 9.8 & 11.3 \\
\hline \noalign {\smallskip}
\end{tabular}
\tablefoot{
$v_{\rm turb}$ is fixed to 5000 km s$^{-1}$ for both UFO tables. $v_{\rm out}$ is derived from the best fit $z$ values, see text for details.}
\vspace{-0.3cm}
\end{table}

We fit UFO$_1$ simultaneously in the full band joint fit, and UFO$_2$ only to the 2024 data set, leaving the column density $N_{\rm H}$, ionization parameter $\log \xi$, and absorber redshift $z$ free to vary for both tables. \autoref{fig:contours} shows the 68, 90, and 99\% confidence contours in the log($\nh$)--log($\xi$) plane for each component, while \autoref{tab:xstar_params} reports the best-fit values. 

For both \xstar tables, the column density is constrained to be up to $\nh=1-2\times10^{24}$ $\cmsq$, i.e. in the Compton thick regime, but with large uncertainties toward lower values, while the ionization parameter is constrained only as a lower limit (log$(\xi)>3.5-4$ erg cm s$^{-1}$) as it reaches the upper boundary explored by the adopted \xstar table, (log$(\xi)=6$ erg cm s$^{-1}$). The improvement in the fit statistic is comparable to that obtained with a single Gaussian line for each UFO (\Cref{tab:ufo_gaussians}), indicating that each \xstar component produces a single blended Fe XXV/XXVI absorption feature that essentially fits each of the observed absorption troughs.

For both tables, log($\nh$) and log($\xi$) are strongly correlated (\autoref{fig:contours}), as higher ionization states produce weaker absorption troughs that, in turn, can accommodate larger columns to fit the same observed features.
This is commonly seen in low-resolution spectra, where the different Fe ions features are blended, and it is not possible to fit them separately \citep{Matzeu16,Lanzuisi24}. For comparison, high-resolution data allow for a significantly more accurate derivation of log($\xi$) and hence log($\nh$) in addition to revealing multiple kinematic components in the wind, as in the cases of PDS456 \citep{xrismpds456}, PG1211+143 \citep{Mizumoto25}, IRAS05189–2524 \citep{Noda25}
and NGC4151 \citep{Xiang25}.

We note that the confidence intervals reported in \autoref{tab:xstar_params} are the 1D 90\% errors for one interesting parameter ($\Delta\chi^2 = 2.706$), while the contour plots in \autoref{fig:contours} show the 2D 90\% joint confidence regions for two interesting parameters ($\Delta\chi^2 = 4.605$, \citealp{Avni76}), which naturally encompass a broader range along each axis, particularly for correlated parameters.
The reported results refer to the fit where the model is reproduced with PEX. However, consistent results are obtained in the other two cases, REFL and CREFL.

Finally, \autoref{fig:zcontours} shows the redshift contours from the same fit, with the relative 68, 90, 99\% confidence levels. The $z$ constraints are converted to $v_{\rm out}$ values using the Doppler formula 
$1+z_a = ((1+\beta)/(1-\beta))^{1/2}$ 
where $\beta=v/c$ and the relations between intrinsic absorber redshift $z_a$, observed redshift $z_o$ and source cosmological redshift $z_c$ is given by $(1 + z_o) = (1 + z_a)(1 + z_c)$ \citep[e.g.,][]{Tombesi11}.
From the redshift constraints we derive $\vout=0.09^{+0.04}_{-0.02}c$ for UFO$_1$ and $\vout=0.33^{+0.02}_{-0.02}c$ for UFO$_2$.


\subsection{Energetics}
\label{sec:energetics}

To compute mass outflow rates and energetics for the two UFOs, the standard approach is based on deriving the absorber location assuming it is bounded between the escape-radius lower limit and the photoionization upper limit,
\vspace{-0.1cm}
\begin{equation}
r_{\rm min} = \frac{2G M_{\rm BH}}{v_{\rm out}^2}, \qquad
r_{\rm max} = \frac{L_{\rm ion}}{\xi N_{\rm H}}
\label{eq.1}
\end{equation}
with $L_{\rm ion}$ integrated over 1--1000 Ryd.

In the case of WISSH 13, the small errors on $\vout$ for UFO$_1$, imply that $r_{\rm min}$ is tightly constrained at 
$r_{\rm min}=9.2_{-1.0}^{+1.2}$ $R_{\rm S}$ (Schwarzschild radii) for UFO$_2$, while it covers a larger range for UFO$_1$: $r_{\rm min}=123_{-64}^{+84}$ $R_{\rm S}$. On the other hand, given the poorer constraints on log($\nh$) and log($\xi$), $r_{\rm max}$ is loosely constrained to be 
larger than a few hundred $R_{\rm S}$ for UFO$_2$ and a few thousand for UFO$_1$.
Therefore, in the following, as a conservative and often adopted approach, mass outflow rates and energetics are evaluated at $r_{\min}$ and treated as lower limits.

We note that the escape-velocity assumption for  $r_{\min}$ is clearly an oversimplification; however, the unresolved Fe K absorption and the absence of a P-Cygni profile prevent us from applying self-consistent disc-wind models \citep[e.g.,][]{Matzeu22a,Luminari24} that can directly constrain the launching radius, as has been done in higher-S/N nearby cases 
\citep[e.g.,][]{Nardini15,Reeves24}.

We account for the velocity–dependent reduction of the effective opacity at relativistic speeds, following the recipes in \citealp{Luminari20,Luminari24}, i.e., a factor of $\sim1.2$ at  $\vout=0.09$c and $\sim2.0$ at $\vout=0.33$c to recover the intrinsic $\nh$ from the observed one. 
We adopt the outflow mass rate formula in \citealt{NardiniZubovas}:
\vspace{-0.1cm}
\begin{equation}
\dot{M}_{\rm out}=\Omega~ N_H~ m_p~\vout~r 
\label{eq.2}
\end{equation}
where $\Omega$ is the global coverage fraction set to 0.5, $m_p$ is the proton mass, and $r$ is the radius adopted for the wind, which yields results equivalent to other widely used wind-mass formulae in the literature \citep[e.g.][]{Tombesi11,Krongold07}.

Using the quantities reported in \autoref{tab:xstar_params} and taking into account errors on $\nh$ and $\vout$, 
we derive a minimum mass outflow rate of
$\mout=21_{-19}^{+28} \ M_{\odot}/\mathrm{yr}~(\sim0.15~\dot{M}_{\rm acc})$ for UFO$_1$ and 
$\mout=24_{-20}^{+16} \ M_{\odot}/\mathrm{yr}~(\sim0.17~\dot{M}_{\rm acc})$ for UFO$_2$.
We note that, since log($\xi$) is constrained only as a lower limit for both components, the strong log($\nh$)-log($\xi$) degeneracy implies that the true column densities could be systematically lower than the best-fit values reported in \autoref{tab:energetics}, which would in turn reduce the derived mass outflow rates beyond the statistical uncertainties quoted above. We also note that, given the definitions in \autoref{eq.1} and \autoref{eq.2}, faster winds have significantly smaller radii (by a factor of 10 going from $\vout=0.1$ to $0.3c$) and the mass outflow rates effectively scale as $1/\vout$. Taking into account relativistic correction and the different $\nh$ values, the two winds end up having comparable mass outflow rates.

\begin{table}[t]
\caption{UFO energetics for WISSH13.}
\label{tab:energetics}
\centering
\begin{tabular}{lcc}
\hline\hline
Parameter & UFO$_1$ & UFO$_2$ \\
\hline
$r_{\rm min}$ ($R_{\rm S}$) & $123^{+80}_{-64}$ & $9.2^{+1.2}_{-1.0}$ \\
$\dot{M}_{\rm out}$ ($M_\odot\,{\rm yr}^{-1}$) & $21^{+28}_{-19}$ & $24^{+16}_{-20}$ \\
$\dot{M}_{\rm out}/\dot{M}_{\rm acc}$ & $\sim 0.15$ & $\sim 0.17$ \\
$\dot{E}_{\rm kin}$ (${\rm erg\,s}^{-1}$) & $5^{+18}_{-4} \times 10^{45}$ & $8^{+7}_{-7} \times 10^{46}$ \\
$\dot{E}_{\rm kin}/L_{\rm bol}$ & $\sim 0.006$ & $\sim 0.1$ \\
$\dot{P}_{\rm out}$ (dyn) & $3.5^{+3.4}_{-3.2} \times 10^{36}$ & $1.6^{+1.2}_{-1.3} \times 10^{37}$ \\
$\dot{P}_{\rm out}/(L_{\rm bol}/c)$ & $\sim 0.14$ & $\sim 0.6$ \\
\hline
\end{tabular}
\tablefoot{$R_{\rm S}$ stands for Schwarzschild radii. Energetics are computed at $r_{\rm min}$ and should be considered lower limits. Column densities are corrected for relativistic effects. A global covering fraction $\Omega = 0.5$ is assumed.}
\end{table}

Finally, we adopt the relativistic expressions for momentum and kinetic power:
\vspace{-0.1cm}
\begin{equation}
\dot{P}_{\rm out} = \gamma\dot{M}_{\rm out}v_{\rm out}, \qquad
\dot{E}_{\rm kin} = (\gamma-1)\dot{M}_{\rm out} c^2
\end{equation}
where $\gamma = (1-v_{\rm out}^2/c^2)^{-1/2}$.
The derived quantities are summarized in \autoref{tab:energetics}.

Figure~\ref{fig:energetics} compares $\mout$ (left) and $\ekin$ (right) of WISSH13 UFO$_1$ and UFO$_2$, as a function of $\lbol$ compared with literature samples: local Seyferts from \cite{Tombesi10}, intermediate redshift QSOs from the SUBWAYS sample \citep{Matzeu23}, and high-z, mostly lensed QSOs from \cite{Chartas21}. The energetics for the \citealt{Chartas21} sample have been recomputed in \cite{Gianolli24}, using the same \xstar table adopted in \citealt{Matzeu23}, and assuming $r_{\ min}$ as radius, and should therefore be considered lower limits.
We also added three well studied low-z QSOs: PDS~456 \citep{Nardini15}, IRASF~11119 \citep{NardiniZubovas,Lanzuisi24} and Mrk~231 \citep{Feruglio15}. 

The values derived above place UFO$_1$ and UFO$_2$ in WISSH13 among the most massive and powerful UFOs known, in terms of absolute mass outflow rate and kinetic energy (UFO$_2$ in particular). 
However, given the extreme luminosity and accretion rate of WISSH13, when normalized to the QSO's $\macc$ and $\lbol$, they follow the relations between wind and accretion energetics observed over many orders of magnitude from the local Seyferts to the most powerful QSOs: over five orders of magnitude in $\macc$ and six in $\lbol$, UFOs in bright QSOs ($\lbol>10^{46} \ergs$) seem to consistently carry {\it at least} $\sim10\%$ of the AGN accretion power, while lower-luminosity AGN span a wider range, even going down to $<1\%$.



\begin{figure}[t]
\centering
    \includegraphics[width=0.45\textwidth]{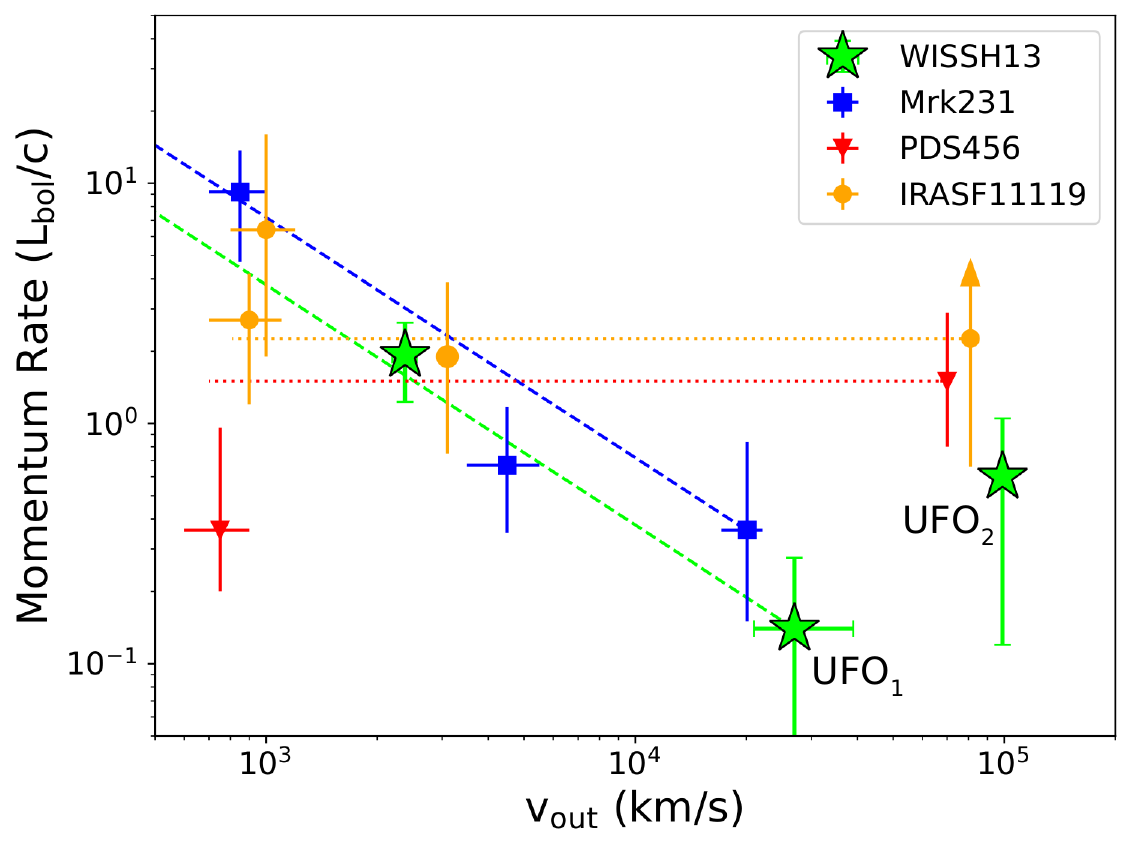}
\caption{Momentum rate vs. velocity of different outflow phases for WISSH13 (green stars), Mrk~231 (blue squares, \citealt{Feruglio15}), PDS~456 (red triangles, \citealt{Bischetti19PDS}) and IRASF~11119 (orange cirlces, \citealt{Lanzuisi24}). 
Data points at $\vout<10^4$ km/s refer to either molecular outflows (CO or OH detection) or to [OIII] and UV winds. Data points at $\vout<10^4$ km/s refer to UFOs. 
Dotted horizontal lines show the evolution of a momentum-driven outflow in the case of PDS~456 and  IRASF~11119,
while dashed lines show the momentum boost 
expected for an energy-driven outflow in the case of Mrk~231 and WISSH13, if UFO$_1$ is considered.}
\label{fig:multi}
\vspace{-0.5cm}
\end{figure}

Finally, in \autoref{fig:multi} we place WISSH13 in the multi–phase wind expansion context, where large-scale (molecular, ionized) outflow momentum rates are compared with those of nuclear outflows. 
A purely momentum-driven expansion implies momentum conservation over the wind expansion (dotted horizontal line), while in a purely energy-driven expansion, the wind-shocked bubble expands adiabatically, and should acquire a momentum boost inversely proportional to the difference in velocity ($P_{out} \propto v^{-1}$ dashed line; \citealt{Zubova12,Costa14}).

We compare WISSH13 with three local QSOs/ULIRGS systems: Mrk~231,  PDS~456, and IRASF~11119. 
These three systems span the full range of observed wind propagation mechanisms, from momentum-conserving (IRASF 11119, \citealt{Veilleux17,Lanzuisi24}) to energy-driven (Mrk 231, \citealt{Feruglio15}) expansion, with PDS 456, dubbed "the gentle monster" \citep{Bischetti19PDS}, showing a large-scale outflow momentum well below the well-studied, strong and persistent nuclear wind \citep{Nardini15,Matzeu16}.

WISSH13 UFO$_2$ appears among the fastest outflows for which such a comparison can be performed
\footnote{With the exception of APM~08279+5255, where large uncertainties on the lensing factor greatly enhance uncertainties on wind energetics \citep{Feruglio17}}, but its momentum seems disconnected from that of the large–scale ionized outflow from [OIII] emission \citep{Bischetti17}. 
Alternatively, the two winds can be reconciled under an energy-driven expansion if the coupling between the two is small, or if a significant fraction of the large-scale wind momentum is missed by only looking at the ionized phase (see \Cref{sec:twophasewind}).
UFO$_1$, on the other hand, seems to be in line with an energy-driven expansion, taking the large–scale outflow energetics at face value.

\section{Discussion and Conclusion}
\label{sec:discussion}

\subsection{Accretion/ejection interplay}
\label{sec:accretion}

The continuum of WISSH13 is best described by a soft intrinsic spectrum with strong reflection and a notably low high-energy cutoff, corresponding to an unusually cold and moderately thick corona. The inferred electron temperature of $kT_{\rm e} \sim 20$~keV and optical depth of order unity place WISSH13 among the coldest coronae measured so far for luminous quasars, while its Eddington ratio $\lambda_{\rm Edd} \sim 3$ clearly places it in the highly accreting regime. This behavior is consistent with the emerging trend of observationally colder coronae at higher luminosities and accretion rates \citep{Peluso26sub,Zhao2026} and with the idea that an increased UV- soft X-ray photon flux from the disk may efficiently cool the corona via Compton scattering, driving it toward a warm, optically thick state. 
This picture is further supported by the recent findings of \citet{Chen25} based on stacked eROSITA spectra of $\sim$4000 type-1 AGN, who showed that the hot corona radius systematically decreases with increasing $\lambda_{\rm Edd}$ while remaining independent of $M_{\rm BH}$, consistent with the moderately thick and cold corona inferred for WISSH13 at $\lambda_{\rm Edd} \sim 3$.

Recent theoretical frameworks by \citet{Inayoshi24} and \citet{Madau24} present two distinct physical scenarios for the X-ray properties of super-Eddington accreting SMBHs like WISSH13. Both models predict the formation of cold coronae coupled to radiation-driven outflows, qualitatively consistent with the low electron temperature ($kT_{\rm e} \sim 20$~keV) observed in WISSH13. However, reproducing the specific spectral slope of $\Gamma \sim 2$, rather than the extremely soft ($\Gamma > 3$) spectra typical of standard super-Eddington simulations, requires a specific configuration that distinguishes the two models. The \citet{Madau24} model explicitly assumes an optically thin corona ($\tau = 0.5$) embedded deep within a reflective funnel. In this geometry, the corona is flooded by soft photons from the funnel walls, which cools the plasma so efficiently that the resulting spectrum is typically steep ($\Gamma \gtrsim 3$). Consequently, this model, in this simplified configuration, struggles to reproduce the $\Gamma \sim 2$ and moderate optical depth ($\tau \sim 2$) measured in WISSH13.

In contrast, the dynamic wind-corona model of \citet{Inayoshi24} may provide a viable solution if the wind properties deviate from standard simulation predictions: super-Eddington simulations typically predict massive outflows comparable to the accretion rate \citep{Hu22,Jiang19}, leading to extremely high optical depths ($\tau_{\rm es} \gg 10$) that would suppress the hard X-ray continuum entirely. However, if the wind is not as massive as these theoretical models predict (a ``starved'' wind), as suggested by our rough estimates on $\dot{M}_{\rm out}$, the resulting corona would naturally have a significantly lower optical depth ($\tau_{\rm es} \sim 1-2$). 

In this scenario, the wind opening angle is an adjustable parameter that can be tuned to match the observed $\tau$: for a low mass wind, moderate collimation can increase the effective density, boosting the optical depth, while a wider opening angle can accommodate a more massive wind. 
The WISSH13 could therefore be in a ``sweet spot'': accreting fast enough to drive a relativistic wind ($v_{\rm out} \sim 0.1-0.3c$) and cool the corona, but with a wind density tuned -- by mass loading and/or collimation -- to allow the hard X-ray continuum to emerge. This finding positions WISSH13 as a critical first case for testing these dynamic accretion-ejection models at Cosmic Noon.


\subsection{A multi-component wind}
\label{sec:twophasewind}

The blind line search in the Fe~K band reveals two significant absorption features, consistent with highly ionized gas outflowing at mildly relativistic speeds. Photoionization modeling confirms that the two features are associated with distinct wind signatures, with velocities $\vout\sim0.1c$ and   $\vout\sim0.3c$, both with high ionization and large column densities. Apart from the archetypal case of the lensed QSO APM~08279, this is the highest-redshift UFO detected to date. At the same time, the equivalent width of the absorption lines in WISSH13 lies at the upper envelope of the distributions for local Seyferts and at the lower end of those reported so far at high redshift. This indicates that long, dedicated spectroscopic campaigns on luminous high-$z$ quasars allow us to move beyond the most extreme UFO detections and to start probing the bulk of the UFO population at Cosmic Noon, with properties more closely resembling those of low-$z$ AGN.

The detection of two distinct velocity components ($\sim 0.1c$ and $\sim 0.3c$) with different variability patterns suggests a complex, stratified outflow. We can quantify the expected variability behavior using the observability 
framework of \citet{KingPounds15}. They assume that UFOs are single thin shells launched instantaneously in each UFO episode, rather than continuous winds. The visibility window of a single shell then depends on its velocity and column density through their Equation~28:
\begin{equation}
t_{\rm off} = \frac{GM_{\rm BH}}{\Omega\,v_{\rm out}^3\,N_H\,\sigma_T} 
\simeq \frac{3\,M_7}{\Omega\,v_{0.1}^3\,N_{22}} \;\text{month},
\end{equation}
where $\Omega$ is the covering factor, $v_{0.1} = v_{\rm out}/0.1c$, 
$N_{22} = N_H/10^{22}$\,cm$^{-2}$, and $M_7 = M_{\rm BH}/10^7\,M_\odot$.
For WISSH13  we obtain $t_{\rm off} \approx 2$~yr (rest-frame) for UFO$_1$ and 
$t_{\rm off} \approx 8$~days for UFO$_2$.

The large $t_{\rm off}$ for UFO$_1$  implies that a single shell remains detectable for $\lesssim 2$~yr after launch. Since UFO$_1$ is consistently detected across the two epochs (separated by $\sim 1.6$~yr rest-frame), the wind recurrence timescale must be $\lesssim t_{\rm off} \sim2$~yr, ensuring that a detectable shell is always present along the line of sight. This naturally explains the consistent detection across the two epochs and supports its interpretation as a persistent, frequently relaunched wind streamline.
In contrast, the very short $t_{\rm off}\sim8$~days for UFO$_2$ implies that its shell disperses below detectability within days of launch. As argued by \citet{KingPounds15}, the likelihood of detecting a UFO so shortly after launch is quite low. Therefore, the lack of detection in 2017 can be easily explained by the low probability of observing such a short-lived feature in any given observation. 

This scenario is consistent with an episodic, transient origin for UFO$_2$, possibly triggered by magnetic reconnection events or flares near the inner disk surface, 
as recently observed by XRISM in NGC~3783 \citep{Gu25}. Indeed, in the context of MHD wind models \citep{Fukumura10,Fukumura17}, the existence of a faster ``spine'' launched from the innermost regions of the accretion disk, surrounded by a slower ``sheath'' launched from larger radii, is predicted. The spine is highly sensitive to short-term fluctuations in the magnetic field configuration and is therefore expected to vary on timescales comparable to $t_{\rm flow}$.

Interestingly, the two-velocity structure observed in WISSH13 closely resembles the ``UFO forest'' recently resolved by XRISM/Resolve in PG~1211+143 \citep{Mizumoto25} and PDS~456 \citep{xrismpds456}, where slower ($v \sim 0.1c$) winds from the outer accretion disk coexist with faster ($v \sim 0.3$--$0.4c$) winds from the inner slim-disk region.
A similar behavior is reported in \citet{Bertola20} for the Einstein Cross (the lensed QSO Q2237+030 at z=1.695), where the low velocity UFO is recovered in all observations and through stacking, while the high velocity one is only visible in some epochs.
This emerging picture of a stratified, multi-velocity outflow as a common feature of near- and Super-Eddington AGN provides a natural physical framework for the UFO$_{1,2}$ detection in WISSH13.
In this scenario, UFO$_1$ may be the primary long-term driver of feedback, while UFO$_2$ has a low duty cycle and therefore limited impact on its surroundings.
Consequently, when evaluating the momentum boost diagram (\autoref{fig:multi}), it is more useful to compare the momentum of the stable UFO$_1$ to the kpc-scale ionized outflow. In this case, WISSH13 aligns well with an energy-conserving expansion scenario ($P_{out} \propto v^{-1}$), where the nuclear wind powers the kpc-scale [OIII] outflow via adiabatic expansion \citep{Zubova12, Costa14}, while UFO$_2$ energetics are disconnected/weakly coupled with those of the large-scale outflow.

Alternatively, UFO$_1$ could be interpreted as a failed wind \citep{Proga04,Giustini19}. In this case, the gas is launched at a radius smaller than the escape radius for its observed velocity, causing it to stall and fall back toward the disk before escaping the nuclear region. In the slim-disk framework, radiation-driven winds are expected to be launched from the inner funnel at radii of order tens of $R_{\rm S}$ \citep{Hu22,Jiang19}, well below $r_{\rm min}$ for UFO${_1}$, lending quantitative support to the failed-wind hypothesis for this component. 
If UFO$_1$ is indeed a failed outflow, its energetics (calculated assuming the wrong radius) are overestimated and, in any case, irrelevant for the momentum boost diagram, as it cannot interact with the galactic-scale feedback. Under this scenario, UFO$_2$ is the only potential driver for the large-scale outflow, and the comparison with the large-scale outflow in \autoref{fig:multi} would be more in agreement with a small coupling between the two phases or a low-duty-cycle nuclear wind, in agreement with its transient nature. 

Indeed, one proposed explanation for the heterogeneous distribution of momentum boosts in the literature invokes the time-variable nature of AGN accretion \citep{NardiniZubovas,Lanzuisi24}: the characteristic timescale for significant accretion rate variations ($\sim 10^{4}$–$10^{5}$~yr; e.g., \citealt{Schawinski2015}) is much shorter than the propagation timescale of the wind to kpc scales ($\sim 10^{6}$–$10^{7}$~yr; e.g., \citealt{ZubovasNardini}). As a consequence, the nuclear momentum flux fluctuates rapidly as the AGN flickers, while the large-scale side of Figure 8 records the time-averaged effect of many outflow episodes.
We also note that the large-scale momentum rates derived from the [O III] ionized phase carry large uncertainties, as the ionized gas may trace only a fraction of the total outflowing mass, with the bulk likely residing in the molecular phase \citep[e.g.,][]{Bischetti19,Travascio24}. A robust comparison of nuclear and galaxy-scale wind energetics, therefore, requires spatially resolved molecular gas observations.

Comparing the observed 1.4 GHz rest-frame luminosity with the radio power expected from wind-driven shocks
using equation 32 from \citet{Nims15}, and adopting as kinetic luminosities $L_{\rm kin}=5\times10^{-4}~\lbol$
from \citet{Bruni19}, the two are consistent with a wind coupling efficiency of $\sim6\%$, just below the $7\%$ threshold above which winds alone cannot account for the observed radio emission \citep{Sun17}. However, we still cannot exclude the presence of a jet, possibly linked to the inner and faster UFO component (Fanelli et al. in prep.). Future VLBI studies will help us characterize the pc-scale morphology and properties of QSOs, thereby distinguishing between wind and jet contributions.


\subsection{Conclusions}

In summary, WISSH13 is a laboratory for accretion and ejection under extreme conditions at $z\sim3$, and it shows the potential of the recently approved WISSHFUL program to probe, in a systematic way, the interplay between accretion properties and nuclear outflows in the most luminous quasars at Cosmic Noon. Its cold, moderately thick corona, and the powerful variable nuclear wind may be examples of the physical conditions envisaged by recent models of super-Eddington accretion and radiation-driven outflows. Extending this type of analysis to the full WISSHFUL sample, in conjunction with other samples from the local Universe to the epoch of reionization, will provide a unique opportunity to understand how SMBH growth, coronal properties, and multiphase feedback are linked throughout the entire SMBH/host galaxy assembly history.

Within a self-regulated 'BH-weather' cycle, turbulence-driven condensation and chaotic inflow lead to intermittent fueling episodes to which the corona/disk can respond rapidly \citep{Gaspari13_cca,Gaspari17_cca,Wittor20}. The emergence of a fast ionized outflow in 2024, absent in 2017, is qualitatively consistent with a duty cycle in which the wind is launched only during phases of enhanced accretion, motivating modeling strategies that allow the absorber strength to vary between epochs.

Advancing our understanding of nuclear wind properties requires new approaches: high-resolution X-ray spectroscopy with XRISM \citep{Xrism} is proving crucial to resolve the velocity structure and ionization stratification of the wind and break the $N_{\rm H}$–~$\xi$ degeneracy that dominates the UFO energetics error budget. Results on PDS~456 \citep{xrismpds456} and NGC~3783 \citep{Gu25} already demonstrate the power of high spectral resolution to reveal multiple kinematic components and track wind acceleration in real time. For fainter high-z QSOs, such as those addressed in this work, we are currently limited to CCD-like resolution. Only the future ESA L mission NewAthena \citep{Cruise2025} will allow us to extend these studies from local Seyferts to highly accreting QSOs at Cosmic Noon and beyond, where accretion and ejection conditions are far more extreme — as demonstrated here for WISSH13 — and AGN feedback potentially more impactful.


\begin{acknowledgements} 
The authors acknowledge support from the INAF Large Program ``DELUX'' of the “Ricerca Fondamentale 2024” INAF program.
EB acknowledges the support of the INAF GO grant ``A JWST/MIRI MIRACLE: Mid-IR Activity of Circumnuclear Line Emission'' and of the “Ricerca Fondamentale 2024” INAF program (mini-grant 1.05.24.07.01). 
T.M. acknowledges financial support from the JSPS KAKENHI grant No. 25K01038 "Comprehensive study of the feedback efficiency of AGN outflows to host galaxies.
M.G. acknowledges support from the ERC Consolidator Grant \textit{BlackHoleWeather} (101086804).
FS acknowledges financial support from the PRIN MUR 2022 2022TKPB2P - BIG-z, “Ricerca Fondamentale 2023” INAF grant 1.05.23.03.04 ``ARCHIE ARchive Cosmic HI \& ISM  Evolution'', “Ricerca Fondamentale 2024” INAF grant 1.05.24.07.01 "ECHOS", Bando Finanziamento ASI CI-UCO-DSR-2022-43 CUP:C93C25004260005 project ``IBISCO: feedback and obscuration in local AGN''.

\end{acknowledgements}

\thispagestyle{empty}

\bibliographystyle{aa}
\bibliography{references}

\begin{appendix}

\section{Monte Carlo Simulations}
\label{app:mc}

To assess the statistical significance of the absorption features at 7.6 keV and 9.7 keV, we performed extensive Monte Carlo simulations as described in \citet{Lanzuisi24} and \citet{Matzeu23}. We simulated 10000 sets of pn and MOS1+2 spectra for the four exposure times listed in \autoref{tab:dataset_info}. As a baseline model, we adopted the best-fit continuum model without any lines, then refitted it simultaneously to each set of spectra, added a narrow Gaussian line in emission or absorption (normalization free to vary in the [-1,+1] range), and searched for the most significant spurious feature for each set of spectra across the full $5-12$ keV band.

\begin{figure}[t]
    \centering
    \includegraphics[width=0.45\textwidth]{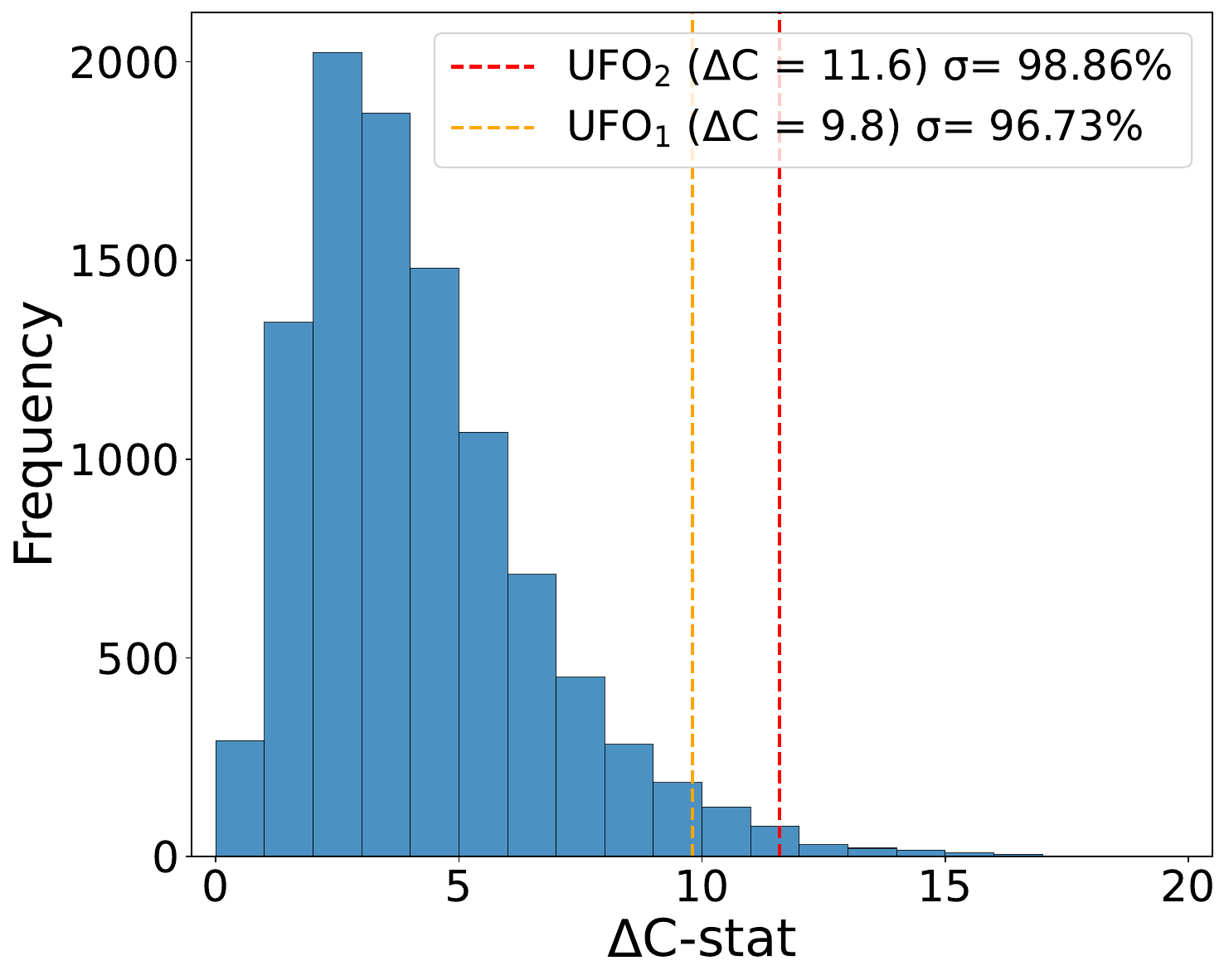}
    \caption{Distribution of $\Delta C$ from the MonteCarlo simulations. The observed $\Delta C$ for UFO$_1$ and UFO$_2$ are shown with orange and red vertical lines, respectively. The resulting significances are $96.73\%$ and 
    $98.86\%$ respectively.
    }
    \label{fig:DeltaC}
\end{figure}

The resulting distribution of $\Delta C$ values, shown in Figure~\ref{fig:DeltaC}, is then compared to the observed $\Delta C$ from UFO$_1$ and UFO$_2$ in the actual data. The features at 7.6 and 9.7 keV are at $96.73\%$ and $98.86\%$ confidence thresholds, respectively, confirming their high statistical significance.

These confidence levels are larger than the typical threshold adopted in systematic studies done at lower redshift, such as those reported in \citet{Tombesi10} and \citet{Matzeu23}, where a confidence level threshold of $95\%$ from MonteCarlo simulations is adopted. These results support the interpretation of these features as signatures of a powerful ultra-fast outflow, consistent with the detection of winds in other hyper-luminous QSOs at high redshift.

\section{Merged 2024 spectrum}

\begin{figure}[t]
\centering
\includegraphics[width=0.90\linewidth]{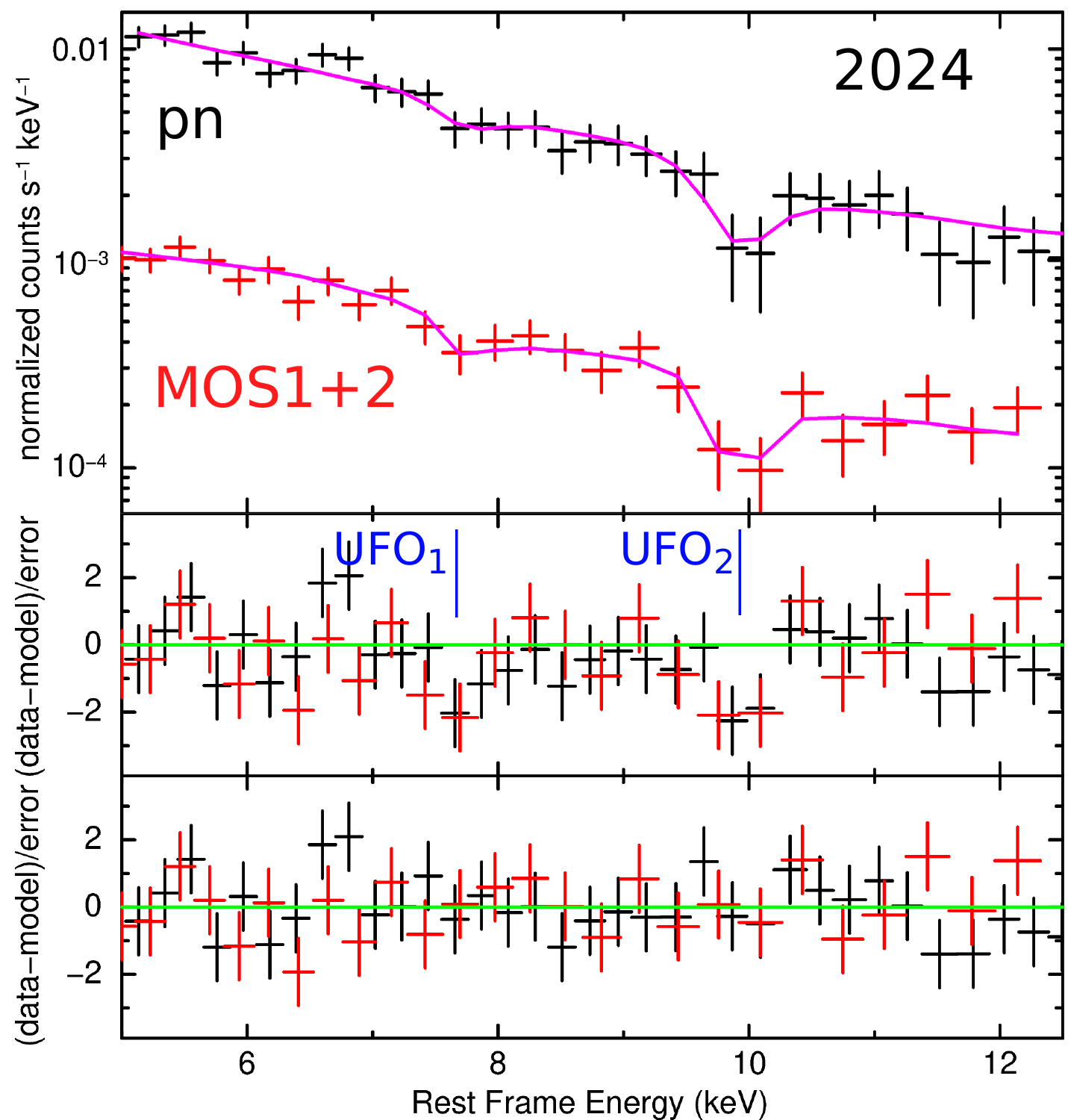}
\caption{2024 merged spectra: EPIC-pn (black) and MOS1+MOS2 (red). The best-fit model, including two absorption Gaussian lines, is shown in magenta. Middle: residuals without the two Gaussian absorption lines. Bottom: residuals with the two Gaussian absorption lines at the best-fit energy. The stack uses strictly flare-cleaned time intervals over the 2024 data set to ensure comparable background levels among the stacked spectra, and therefore has lower exposure time and counts (by 50\%) than the full, jointly fitted 2024 spectra adopted in \Cref{sec:analysis}.}
\label{fig:merged}
\end{figure}

For visualization purposes only, we show in \Cref{fig:merged} the stacked spectra obtained by co-adding all 2024 EPIC-pn (black) and all MOS1+MOS2 data (red). The middle panel of \Cref{fig:merged} shows the residuals without including two narrow Gaussian absorption lines, while the bottom panel shows residuals with the lines at the best-fit energies. 

This stack is intended for illustrative purposes only: to avoid mixing different background levels and producing biased merged spectra, we must fully erase soft-proton flare intervals in all exposures before stacking. This requirement reduces the usable exposure time by $\sim50\%$ with respect to the optimized reduction adopted in \Cref{sec:reduction}, significantly degrading the total number of counts and the S/N of the merged 2024 spectrum compared to the simultaneous joint fit on the individual 2024 spectra presented in \Cref{sec:analysis}.

The 2017 data set, which makes the strongest contribution to the $\sim8$ keV rest-frame feature associated with UFO$_1$, is not included in the stack. As a result, UFO$_1$ is only marginally recovered in the stacked representation, although it is robust in the joint fit. On the other hand, UFO$_2$ is clearly visible even in the stacked, lower-quality spectrum.
We stress that pn and MOS1+2 show similar residuals at the same energy for both UFOs, further reinforcing our confidence in their detection.  

\end{appendix}

\end{document}